\documentclass[10pt,journal,compsoc]{IEEEtran}
\ifCLASSOPTIONcompsoc
  \usepackage[nocompress]{cite}
\else
  \usepackage{cite}
\fi
\usepackage{graphicx}
\ifCLASSINFOpdf
\else
\fi
\usepackage{caption}
\usepackage{subcaption}

\usepackage{url}

\usepackage{comment}
\usepackage{xcolor}

\begin{document}
%
\title{Privacy Engineering in the Wild: Understanding the Practitioners' Mindset, Organisational Aspects, and Current Practices}

%
%
%
%

\author{Leonardo~Horn~Iwaya,~\IEEEmembership{Member,~IEEE,}
        Muhammad~Ali~Babar,~%
        and~Awais~Rashid,~\IEEEmembership{Member,~IEEE}
\IEEEcompsocitemizethanks{\IEEEcompsocthanksitem L. H. Iwaya is with the Centre for Research on Engineering Software Technologies, the University of Adelaide, Adelaide, SA, Australia, 5005; the Cyber Security Cooperative Research Centre (CSCRC), Australia; and, the Privacy and Security (PriSec) research group, Department of Mathematics and Computer Science, Karlstad University, Sweden.
\protect\\
E-mail: leonardo.iwaya@kau.se
\IEEEcompsocthanksitem Muhammad Ali Babar is with the Centre for Research on Engineering Software Technologies, the University of Adelaide, Adelaide, SA, Australia, 5005; and, the Cyber Security Cooperative Research Centre (CSCRC), Australia.
\IEEEcompsocthanksitem A. Rashid is with the Bristol Cyber Security Group, Department of Computer Science, University of Bristol, United Kingdom; and, REPHRAIN: National Research Centre on Privacy, Harm Reduction and Adversarial Influence Online, United Kingdom.
\protect\\
E-mail: awais.rashid@bristol.ac.uk}
\thanks{Manuscript received Month Day, 2023; revised Month Day, 2023.}}

%
%

\markboth{IEEE Journal Template,~Vol.~x, No.~x, Month~2023}%
{Iwaya \MakeLowercase{\textit{et al.}}: Privacy Engineering in the Wild}
%



\IEEEtitleabstractindextext{%
\begin{abstract}
Privacy engineering, as an emerging field of research and practice, comprises the technical capabilities and management processes needed to implement, deploy, and operate privacy features and controls in working systems. 
For that, software practitioners and other stakeholders in software companies need to work cooperatively toward building privacy-preserving businesses and engineering solutions.
Significant research has been done to understand the software practitioners' perceptions of information privacy, but more emphasis should be given to the uptake of concrete privacy engineering components.
This research delves into the software practitioners' perspectives and mindset, organisational aspects, and current practices on privacy and its engineering processes. 
A total of 30 practitioners from nine countries and backgrounds were interviewed, sharing their experiences and voicing their opinions on a broad range of privacy topics. 
The thematic analysis methodology was adopted to code the interview data qualitatively and construct a rich and nuanced thematic framework. 
As a result, we identified three critical interconnected themes that compose our thematic framework for privacy engineering ``in the wild'': 
(1) personal privacy mindset and stance, categorised into practitioners' privacy knowledge, attitudes and behaviours; 
(2) organisational privacy aspects, such as decision-power and positive and negative examples of privacy climate; and, 
(3) privacy engineering practices, such as procedures and controls concretely used in the industry.
Among the main findings, this study provides many insights about the state-of-the-practice of privacy engineering, pointing to a positive influence of privacy laws (e.g., EU General Data Protection Regulation) on practitioners' behaviours and organisations' cultures.
Aspects such as organisational privacy culture and climate were also confirmed to have a powerful influence on the practitioners' privacy behaviours.
A conducive environment for privacy engineering needs to be created, aligning the privacy values of practitioners and their organisations, with particular attention to the leaders and top management's commitment to privacy.
Organisations can also facilitate education and awareness training for software practitioners on existing privacy engineering theories, methods and tools that have already been proven effective.
\end{abstract}

}

\maketitle

\IEEEdisplaynontitleabstractindextext

%
\IEEEpeerreviewmaketitle

\IEEEraisesectionheading{\section{Introduction}\label{sec:introduction}}

%
%
%
%
\IEEEPARstart{P}{rivacy} has become a significant concern to governments and technology companies worldwide.
As our digital society relentlessly expands, all aspects of people's lives are transformed into pieces of data collected and processed by a multitude of information systems.
Organisations that handle personal data are increasingly expected by their customers, employees, and multiple stakeholders to build systems that respect people's right to privacy \cite{stallings2019information}.
Such demands are accompanied by a new wave of legal privacy frameworks.
Over the last decade, countries from every major region globally have continued to enact data privacy laws, raising to 157 the number of countries with such regulations as of mid-March 2022 \cite{greenleaf2022now}.

The most influential privacy law today, the European Union's General Data Protection Regulation (GDPR) \cite{GDPR2016, greenleaf2022now}, was designed to apply to all types of businesses, from multinationals to micro-enterprises. 
Unlike previous regulations, the EU GDPR can lead to hefty fines, and any non-compliant organisation faces a significant liability regardless of its size.
Serious infringements could result in a fine of up to 20 million euros, or 4\% of the firm's worldwide annual revenue from the preceding financial year, whichever amount is higher (Art. 83 GDPR \cite{GDPR2016}).
Furthermore, considering the GDPR's extraterritorial application, companies outside the EU could also be affected.
For instance, the US big tech Alphabet Inc. (``Google'') could receive a \$4 billion fine in cases of serious infringements \cite{houser2018gdpr}, since it also operates in the EU and processes data of EU citizens.
Given that, it was expected that the GDPR would radically affect American technology companies operating with rather loose privacy restrictions under US law \cite{houser2018gdpr}.
Similar ripple effects of privacy have been observed in many other countries as the GDPR casts its net worldwide.

Given that software engineers are often responsible for designing, managing, testing and deploying a series of technical and organisational privacy controls that need to be put in place to fulfil regulatory compliance, it is imperative that they not only fully appreciate the importance of privacy concerns, but also are knowledgeable about existing privacy laws and standards. However, the interpretation of legal documents tends to be vague, and it is hard for practitioners to translate such legal requirements into concrete engineering practices.

Privacy engineering responds to this gap between law, research and practice. As defined by G{\"u}rses and Del \'{A}lamo, \textit{``[p]rivacy engineering is an emerging research framework that focuses on designing, implementing, adapting, and evaluating theories, methods, techniques, and tools to systematically capture and address privacy issues in the development of sociotechnical systems''} \cite{gurses2016privacy}.
Alternatively, as defined by Stallings \cite{stallings2019information}, privacy engineering \textit{``encompasses the technical capabilities and management processes needed to implement, deploy, and operate privacy features and controls in working systems''}.
Examples of privacy features are privacy notices, consent management platforms, transparency-enhancing tools (e.g., privacy dashboards), data portability, and data deletion mechanisms (``right to be forgotten'').

With that in mind, many studies have addressed privacy through the perspectives of software practitioners, particularly using interview-based studies that allow for deep insights into the practitioner's views and industry practices.
Existing interview studies usually emphasise the practitioner's privacy perceptions \cite{hadar2018privacy, peixoto2020understanding} and personal factors \cite{arizon2021understanding}.
Other studies are more cohesive, focusing on the privacy perspectives of specific roles, such as app developers \cite{balebako2014privacy, li2018coconut}, and senior engineers \cite{bednar2019engineering}.
However, significant research gaps still exist, such as the lack of studies with emphasis on privacy engineering and addressing the practitioner's perspectives in a ``post-GDPR'' world and the uptake of concrete privacy practices in the industry.
Furthermore, the existing studies tend to overly rely on subjects from Western, Educated, Industrialised, Rich and Democratic (WEIRD) societies (i.e., US and Europe), which may result in findings that are not representative of non-WEIRD counterparts \cite{henrich2010most}.

For such reasons, we conducted this empirical study on privacy engineering ``in the wild'', adopting an interview-based study similar to Hadar \textit{et al.} \cite{hadar2018privacy}, which addressed the closely related topic of privacy-by-design.
Our study reached a broad participant base, interviewing 30 practitioners from 29 companies of nine countries and backgrounds.
We were able to include practitioners from several countries, mainly from outside of the North American and European regions where other studies have drawn their data, allowing our findings to be based on a further diversified number of views on privacy engineering.
This differentiation is important, especially now that the global effects of international privacy laws affect several software organisations also in low- and middle-income countries.

The thematic analysis methodology \cite{braun2006using} was utilised for interpreting the interviews' data, following a primarily inductive process.
As a result, this study's main contribution is the creation of a rich and nuanced thematic framework composed of three major themes: (1) the practitioners' mindset, (2) the organisational privacy aspects, and (3) the concrete engineering practices used to address privacy concerns in the industry.
Special attention was given to privacy engineering practices currently implemented, i.e., specific technical and organisational privacy controls, yet also capturing the privacy perceptions of the software practitioners on several aspects.

In a nutshell, the novel findings of our study are:
\begin{itemize}
    \item The practitioners' privacy awareness is increasing, compared to previous studies \cite{hadar2018privacy, bednar2019engineering} that suggested more lack of knowledge and negative experiences. However, they remain mostly unaware of existing standards related to privacy engineering.
    \item The organisation's privacy culture and climate influence the practitioners' behaviours. Situations that contribute to an adverse privacy climate are the superiors' lack of privacy knowledge, the lower priority and value given to privacy, and the lack of incentives for privacy training.
    \item Privacy is mainly addressed on a project basis and in an unsystematic manner. Although practitioners show more awareness about privacy strategies, encryption mechanisms are highly prevalent, while other privacy strategies are less commonplace.
\end{itemize}

We also identified key challenges in privacy engineering that point to potential pathways for future research.
Addressing such challenges requires the involvement of multiple stakeholders (e.g., practitioners, leaders, researchers, and standardisation bodies) in order to improve personal, organisational and engineering practices related to privacy.

\section{Background}
\label{sec:background}
\subsection{Contrasting Security and Privacy}
The term ``privacy'' is often confused as a subset of the broader category of ``security''.
Nonetheless, privacy has gained its own distinct significance and prominence over time.
For this research, it is crucial to clarify the differences between privacy and security.

The security of computing systems aims to protect and safeguard hardware, software, and information, typically focusing on three key properties \cite{bishop2005introduction}: confidentiality, integrity, and availability (CIA).
Confidentiality involves concealing information or resources, integrity involves ensuring the trustworthiness of data resources and preventing improper or unauthorised alterations, and availability relates to the ability to access desired resources or information.

Privacy has, however, further dimensions, as seen in many privacy laws such as the GDPR \cite{GDPR2016}.
These principles, such as lawfulness, consent, purpose binding, data minimisation, and transparency, are disjointed or have little overlap with information security.
Privacy can also be compared to the CIA triad through the proposed additional privacy protection goals of unlinkability, transparency, and intervenability introduced by \cite{hansen2012top,hansen2015protection}.
Unlinkability refers to separating data and processes to ensure that personal data is unlinkable to any other set of personal data outside of the domain.
Transparency refers to adequately and clearly describing personal data processing activities so that the collection, processing, and use of information can be understood and reconstructed at any time.
Intervenability refers to the data subject's ability to interfere with personal data collected or processed.

Therefore, it should be clear that although information privacy is protected through information security measures, privacy cannot be satisfied solely based on managing security \cite{NIST8062}.

\subsection{Privacy Legal Frameworks}
Many privacy laws have been enacted in the last decade, with the EU GDPR being the most prominent.
The GDPR replaced the previous Directive 95/46/EC \cite{GDPR2016} and was designed to: 
(a) harmonise privacy and data protection laws across Europe; 
(b) protect and empower all EU citizens' privacy and data protection; and, 
(c) reshape the way organisations across the region approach privacy and data protection.
The GDPR does not apply only to EU member states but also to organisations (i.e. data controllers and data processors) outside the EU that offer goods and services to, or that monitor, individuals in the EU. 
For this reason, the GDPR casts its net globally. 
Many other countries are updating their privacy laws to guarantee an \emph{adequate level} of data security and protection. 
Such adequacy facilitates cross-border data transfers among international business activities, avoiding penalties and administrative fines. 
Today, the GDPR is considered state-of-the-art in privacy law and impacts organisations worldwide.

Besides the GDPR, two other privacy laws are repeatedly mentioned throughout this study, as reported by many interview participants.
First, the Brazilian \emph{Lei Geral de Proteção de Dados} \cite{LGPD2018} (BR LGPD, General Law for Data Protection), which is significantly akin and draws from the GDPR. 
Second, the Australian Privacy Act 1988 \cite{APA2021} (APA), which was introduced to regulate how Australian Government agencies and organisations with an annual turnover of more than 3 million should promote and protect individuals' privacy.
It is worth noting, however, that according to the EU Commission, both BR LGDP and APA still do not offer adequate levels of data protection if compared to the GDPR.
As of 2021, the EU Commission has so far recognised that Andorra, Argentina, Canada, Faroe Islands, Guernsey, Israel, Isle of Man, Japan, Jersey, New Zealand, Republic of Korea, United Kingdom, Switzerland, and Uruguay as providing adequate protection \cite{eu2021adequacy}.

\subsection{Privacy Engineering}
There are currently many privacy engineering theories, methods, tools, and techniques to systematically capture and address privacy in developing software systems \cite{gurses2016privacy}.
This section aims to introduce a few critical components related to privacy engineering, yet without the intention of providing a broad summary of the topic.
For a more comprehensive view of the topic, we refer readers to other references, such as \cite{gurses2016privacy, stallings2019information}.

We can associate the origin of privacy engineering with the rise of \emph{Privacy-Enhancing Technologies} (PETs) \cite{fischer2017privacy} that were first introduced in the 1980s and have since considerably evolved in computer science.
PETs refer to the use of information and communication technologies (i.e., software or hardware) for protecting informational privacy \cite{van2003handbook}.
This is typically done by (a) minimising the use of personal data that is processed, (b) ensuring information security, and (c) empowering individuals through transparency and more control over their data.
Examples of PETs are communication anonymisers, zero-knowledge proofs, homomorphic encryption, access control mechanisms, transparency-enhancing technologies, privacy policy languages, etc.

However, given that PETs are purely technological artefacts, they cannot address the entire scope of individual privacy \cite{ceross2018rethinking}. 
Informational privacy is, after all, a social construct with social consequences, and thus it cannot be dealt with solely based on technology.
Hence, privacy engineering acknowledges the necessity of thinking about sociotechnical systems.
Several \emph{privacy requirements} can be derived from regulations, but the law tends to be vague and subject to legal interpretation \cite{colesky2019helping}.
Some vital initial contributions in the area are the definition of the \emph{Privacy-by-Design principles} \cite{cavoukian2009privacy} and their integration into concrete software engineering practices \cite{gurses2011engineering}.
Other researchers have also worked in this gap between the law and engineering practices, e.g., by proposing \emph{privacy design strategies} \cite{hoepman2014privacy, colesky2016critical}, which makes more sense to software engineers.
There are also approaches for defining \emph{privacy design patterns} as well as evading the so-called privacy ``dark patterns'' in software development \cite{hafiz2006collection,lenhard2017literature}.

Personal data privacy can also be approached from a risk management perspective.
\emph{Privacy Impact Assessments} (PIAs) are one of the primary methods organisations can use to address privacy risks.
A PIA is a systematic process for evaluating the potential effects on the privacy of a project \cite{clarke2009privacy}.
As a process, PIAs usually start with Data Flow Diagrams (DFDs) for identifying different parties that collect and process personal data.
Based on this high-level understanding, a system can be analysed in terms of the relevant privacy principles (derived from legal frameworks), followed by an in-depth \emph{threat modelling} that comprises the identification of \emph{privacy threats} and assignment of \emph{privacy controls} to avoid, eliminate or mitigate risks.
Privacy controls, in turn, refer to technical and organisational measures, e.g., privacy policies, PETs, and privacy design patterns.

Moreover, standardisation bodies now offer guidance to support privacy engineering in the software development life cycle.
Examples are the ISO/IEC TR 27550:2019 \cite{ISO27550} standard and the NIST Privacy Framework \cite{NISTPF, NIST8062}.
Overall, it is noticeable that the field of privacy engineering started to crystallise around a significant body of knowledge and generally accepted practices \cite{gurses2016privacy}.

\subsection{Organisational Climate and Culture}
\label{sec:org-climate-culture}
Organisational climate and organisational culture are two important overlapping constructs for studying how people perceive, experience and describe their work settings \cite{brown1996new, schneider1983etiology, schneider2013organizational}.
The research on organisational climate can be traced back to the 1950s and 1960s, while the interest in organisational culture spread in the 1970s and 1980s \cite{schneider2017organizational}.

The organisational climate construct is generally defined as \textit{``the shared perceptions of the meaning attached to the policies, practices, and procedures employees experience and the behaviors they observe getting rewarded and that are supported and expected''} \cite{schneider2013organizational} (p. 362). 
Initially, organisational climate was conceptualised as a broad construct (i.e., the whole organisational functioning).
Later, there was a growing interest in an organisational climate with a specific focus or facet.
Examples include customer service climate \cite{bacile2020digital}, safety climate\cite{bhandari2022influence}, initiative climate \cite{vihari2022impact}, learning climate \cite{peng2022learning}, information security climate \cite{dong2021effect}, and privacy climate \cite{hadar2018privacy}.
Most organisational climate research comes from scholars trained in psychological methods, almost entirely using employee survey methods, with surveys targeting people's observable experiences in their work environment \cite{schneider2017organizational}.

Organisational culture has its conceptual and methodological basis in sociology and anthropology and was largely embraced by areas such as organisational studies and organisational psychology.
Organisational culture can be broadly defined \textit{``as the shared basic assumptions, values, and beliefs that characterize a setting and are taught to newcomers as the proper way to think and feel, communicated by the myths and stories people tell about how the organization came to be the way it is as it solved problems associated with external adaptation and internal integration''} \cite{schneider2013organizational} (p. 362).
The methods adopted by early culture researchers were mainly qualitative, stressing the importance of immersion in the studied setting \cite{schneider2017organizational}.

Examples of the prevalent themes within organisational culture research are (1) leadership and (2) national cultures \cite{schneider2013organizational}.
Leaders embed the desired values in their organisations through multiple primary (i.e., resource allocation, rewarding systems and status, selection and promotion strategies) and secondary mechanisms (i.e., organisational systems, procedures, design and structure, rites and rituals, stories, organisational philosophy, creeds, and charters) \cite{schein2010organizational}. 
To understand how and to what extent national culture shapes organisations located in a given nation, Hofstede's work is the most influential \cite{hofstede1980culture, hofstede2011dimensionalizing}.

Security researchers have also adopted the constructs of organisational culture and climate.
The information security culture refers to \textit{``the attitudes, assumptions, beliefs, values and knowledge that employees/stakeholders use to interact with the organisation's systems and procedures at any point in time''} \cite{daveiga2010framework}.
The term security climate has been used to refer to the employee’s perception of the current organisational state concerning information security as evidenced through dealings with internal and external stakeholders \cite{chan2005perceptions}.

For this study, we consider the narrow facet of privacy under the organisational climate theory that is expected to help researchers understand the employees' shared perceptions, i.e., especially when attempting to assess/measure factors via survey instruments quantitatively.

\subsection{Related Work}
A significant number of studies employing distinct research methods have been conducted to understand how software practitioners and organisations handle privacy in working systems.
Survey-based studies are often used to capture the developers' perspectives \cite{bu2020privacy}, perceptions \cite{canedo2020perceptions}, attitudes \cite{ayalon2017developers, spiekermann2018inside, van2019data}, reported behaviour and perceived responsibility \cite{spiekermann2018inside} with regards to privacy.
Other approaches use task-based experiments to grasp the developers' perspectives through the practical exercise of designing a privacy-sensitive system \cite{senarath2018understanding, senarath2018developers, senarath2019will, alhazmi2021listening}.
Scholars have also turned to large software repositories (e.g., GitHub and Stack Overflow), mining posts and question-and-answer websites to interpret how developers talk about privacy \cite{li2021developers, tahaei2022understanding}.
Lastly, secondary research, in the forms of systematic reviews, has also been able to synthesise this growing body of evidence on factors affecting the implementation of privacy and security practices \cite{nurgalieva2021wip} and the conceptualisation of emerging topics such as Organisational Privacy Culture and Climate (OPCC) \cite{iwaya2022organisational}.

In this section, however, we pay particular attention to the existing interview-based research, which is similar to this study.
Interview studies enable researchers to gain insight into practitioners' perceptions, understandings and experiences on privacy, enabling in-depth data collection.
Sometimes these studies investigate the practitioners' privacy perceptions in very particular fields, such as works on privacy-preserving computation techniques \cite{agrawal2021exploring}, aged care monitoring devices\cite{alkhatib2019privacy}, mobile apps for children \cite{alomar2022developers}, and risks in virtual reality systems\cite{adams2018ethics}.
However, other studies address the topic of privacy more broadly, investigating the overall practitioners' understandings and industry practices, as presented in Table \ref{tab:related-work}.
These studies are considered the main related work that has significantly improved the understanding of the topic and motivated this research.
In what follows, we first briefly introduce these studies so that we can later discuss their limitations and the identified research gaps.

\begin{table*}[ht]
\caption{Summary of related work with a focus on interviewing practitioners about privacy in software development. Note: (\textbf{*}) other countries include the United States, Germany, Norway, the United Kingdom, China and Singapore.}
\begin{center}\small
\begin{tabular}{p{0.085\textwidth}p{0.085\textwidth}p{0.15\textwidth}p{0.08\textwidth}p{0.08\textwidth}p{0.37\textwidth}}
\hline
\textbf{Reference} & \textbf{Data} &\textbf{Subjects} & \textbf{Location} & \textbf{Study's Period} & \textbf{Limitations} \\
\hline \hline

Balebako \textit{et al.} (2014) \cite{balebako2014privacy} & Interview / survey & 13/228 -- App developers & United States & 2013 & (i) Limited to app developers. (ii) All participants from the United States. \\

Hadar \textit{et al.} (2018) \cite{hadar2018privacy} & Interview & 27 -- Software developers practising design or architecture &  \textit{(not mentioned)} & 2013-2014 & (i) Interviewees are only developers with alleged roles in design and architecture. (ii) More focus on developers \emph{perceptions}. (iii) Unclear demographics. \\

Li \textit{et al.} (2018) \cite{li2018coconut} & Interview & 9 -- Android software developers & United States  & 2018 & (i) Limited to Android app developers. (ii) More focus on the \emph{difficulties} with privacy. \\

Sirur \textit{et al.} (2018) \cite{sirur2018are} & Interview & 12 -- Senior executives & United Kingdom & \textit{(not mentioned)} & (i) Only UK organisations. (ii) More focus on the organisations' \emph{experiences} and \emph{processes} for implementing GDPR. \\

Bednar \textit{et al.} (2019) \cite{bednar2019engineering} & Interview / survey & 6/124 -- Senior software engineers/software engineers & \textit{(not mentioned)} & \textit{(not mentioned)} & (i) Interviews only  with senior engineers from (ii) globally leading IT corporations and research institutes. (iii) More focus on the \emph{motivation} and \emph{compliance} with privacy regulations.  (iv) Unclear demographics. \\

Ribak (2019) \cite{ribak2019translating} & Interview & 2 -- Software developers in one start-up & Israel & 2017-2018 & (i) Interview instruments not provided. (ii) Only two interviewees from the same company. \\

Peixoto \textit{et al.} (2020) \cite{peixoto2020understanding} & Interview & 13 -- Software developers in six companies & Brazil & 2019 & (i) Interviewees from the same city. (ii) More focus on \emph{personal factors}. \\
 
Tahei \textit{et al.} (2021) \cite{tahaei2021privacy} & Interview & 12 -- Privacy champions & North America, Europe, and Asia & 2020 & (i) Interviews only with privacy champions. (ii) Strategies are mainly discussed in organisational terms, i.e., less emphasises on interviewees' knowledge and technical practices. \\

Arizon-Peretz \textit{et al.} (2021) \cite{arizon2021understanding} & Interview & 27 -- Software practitioners from fourteen companies &  \textit{(not mentioned)} & \textit{(not mentioned)} & (i) More focus on developers’ \emph{perceptions} and \emph{behaviours} and underlying forces affecting them. (ii) Unclear demographics. \\

Dalela \textit{et al.} (2022) \cite{dalela2022study} & Interview / survey & 11/107 -- Software developers and managers & Denmark & 2020 & (i) Not focused on software development. (ii) More focus on the \emph{challenges} faced by Danish companies. (iii) Unclear demographics. \\

\hline \hline
\textbf{This work} & Interview & 30 -- Software practitioners from twenty-nine companies & Sri Lanka, Brazil, Australia, and six other countries\textbf{*} & 2020-2021 & (i) Not all interviewees are senior engineers. (ii) Some participants do not have ample decision-making power over the systems, yet they are able to influence them. (iii) More focus on developers' \emph{knowledge}, \emph{attitudes}, and \emph{behaviours} and reported \emph{practices} related to privacy engineering. \\
\hline
\end{tabular}
\label{tab:related-work}
\end{center}
\end{table*}

\subsubsection{Focus on Mobile App Development}
As shown in Table \ref{tab:related-work}, two studies have specifically investigated the perceptions of \emph{app developers} on privacy \cite{balebako2014privacy, li2018coconut}.
The work of \cite{balebako2014privacy} is one of the first studies on the topic, showing that smaller companies were less likely to demonstrate positive security and privacy behaviours.
Their interview findings also indicated that app developers had little to no formal education on privacy, lacked knowledge about existing regulations, and would usually consult friends and social networks for advice \cite{balebako2014privacy}.
App developers also relied mainly on off-the-shelf 3rd-party security tools (e.g., for encryption and authentication) and did not have as many tools for privacy (i.e., only privacy policy ``generators'') \cite{balebako2014privacy}.
Besides, their findings also suggested that many app developers regarded privacy as low priority and low value, conflicting with monetisation and perceived it as an extra cost \cite{balebako2014privacy}.

Also focusing on app developers, the work of \cite{li2018coconut} looked into the main \emph{difficulties} faced by practitioners concerning privacy.
Their results showed that app developers just partially understood what privacy is and needed to gain knowledge about technical measures that could be implemented. 
Developers also had an inaccurate understanding of the app's behaviours leading to the creation of inappropriate privacy notices \cite{li2018coconut}.
Privacy was also treated as a secondary task, and developers were susceptible to ignoring privacy issues, mainly when dealing with other technical constraints (e.g., Android's permission system is not fine-grained enough) \cite{li2018coconut}.

\subsubsection{Focus on Specific Practitioners Roles}
Some studies investigated the perspectives of practitioners in specific roles with respect to privacy and compliance challenges.
These roles were software architects \cite{hadar2018privacy}, senior executives \cite{sirur2018are}, senior engineers \cite{bednar2019engineering}, and privacy champions \cite{tahaei2021privacy} (see Table \ref{tab:related-work}).

In \cite{sirur2018are}, the researchers focused \emph{senior executives} and their views on compliance challenges with the GDPR.
According to their results, large organisations find the regulations mostly reasonable and doable, while small-to-medium-size organisations generally struggle.
Among the main challenges were the sheer breadth of the regulation, its difficult interpretation, and the problem of mapping out and taking inventory of large and complex networked systems that process personal data.

The study of \cite{hadar2018privacy} provides many insights into the \emph{software architects}' mindsets through in-depth interviews about information privacy and organisational and technical privacy strategies.
In their work, \emph{organisational privacy climate} was found to be a central force influencing the environment and the developers' cognitive factors and behaviour related to privacy.
The study also articulates positive and negative privacy climates in the organisation (e.g., (+) having clear guidelines, (-) a low sense of responsibility).
The authors also point to a misalignment between the organisations' privacy policies and the actual privacy climate among employees.
For instance, there is little to no concern for privacy when designing and developing systems despite normative privacy policies, or there are mismatches between the norms and employees' moral values.

The work of \cite{bednar2019engineering}, focused on \emph{senior engineers}, highlighted that they make very few expressions of responsibility, autonomy, and control over privacy matters.
Senior engineers reported an overall negative experience concerning privacy, considering it a burden and expressing sceptical or pessimistic views, e.g., avoiding responsibility and listing problems with implementing privacy protections.
Many also perceived a lack of social pressure from the general population, decreasing their motivation toward addressing privacy concerns.
There are also conflicts between the legal and engineering worlds.
Cooperation with lawyers was difficult and tiresome, and it took an effort to reach a shared understanding.

Lastly, the study of \cite{tahaei2021privacy} focuses on \emph{privacy champions}, aiming to understand their motivations, strategies and challenges in software teams.
This work found that common barriers to implementing privacy are the negative privacy culture, prioritisation tensions, and limited tool support.
Privacy champions generally use informal communication channels to promote privacy, e.g., perceiving code reviews as more instructive than privacy awareness and training programs.

\subsubsection{Various Software Practitioners}
The work of \cite{ribak2019translating} draws from two interviews with software developers in an Israeli startup, telling a story about the globalisation of a company and the shaping of the developers' notions of information privacy.
As stated by the authors, external privacy regulations marked the transition from a naive and youthful company into a responsible and mature one.
However, important conflicts were still challenging, such as balancing the collection of information for sales and marketing and respecting the user's privacy.

Other researchers have looked into the \emph{personal factors} that influence developers' perceptions, and interpretation of privacy requirements in software development \cite{peixoto2020understanding, peixoto2022perspective}.
Drawing significantly from the work of Hadar \textit{et al.} \cite{hadar2018privacy}, this interview-based study reveals nine personal factors that were found to positively and negatively affect developers \cite{peixoto2020understanding}.
Positive factors were related to previous knowledge about privacy, experience with user control mechanisms, evaluating privacy on a project basis, and making privacy everyone's responsibility in the organisation.
On the other hand, negative personal factors were linked to the confusion between security and privacy concepts, low importance to user data, privileging security over privacy, lack of formal privacy knowledge, and shifting the responsibility to users to be ``proactive'' over their privacy.
The authors highlighted that companies should foster a culture of privacy through general guidelines.

The work of \cite{dalela2022study} also has focused on the challenges related to GDPR compliance in Danish IT companies.
Among the main challenges, they found a misalignment between software developers and management regarding implementing security and privacy measures.
They also noticed an overall difficulty in adapting company practices to comply with the GDPR, especially in combination with other challenges during the pandemic, i.e., remote work and external access to the company's network.

Lastly, also based on the study of Hadar \textit{et al.} \cite{hadar2018privacy}, the work of \cite{arizon2021understanding} proposes using organisational climate theory for attaining a better understanding of developers' privacy perceptions and behaviours and the underlying forces.
Another research aim was to discover the constructs that compose organisational privacy and security climates.
Their findings reveal that software developers receive inconsistent and confusing cues from management and other parties in their work environment.
Privacy is seen as a low priority, leading to perceptions and behaviours that would not comply with existing regulations.
As a result, this study has provided some foundations for developing climate measures to quantify organisational privacy and security climates.

\subsection{Identified Research Gaps}
Prior work has contributed significantly to the overall understanding of the practitioner's mindset and reported practices, but more research is needed to corroborate findings and expand the scope of participants.
The work of Hadar \textit{et al.} \cite{hadar2018privacy} is arguably the most influential on the topic, from which we also borrowed significantly in terms of methodology and interview approach -- further discussed in Section \ref{sec:methodology}.
Nevertheless, we argue that a continuous investigation of the practitioners' privacy perceptions is essential. 
Also, there is a need to investigate more concretely the practitioner's knowledge and uptake of existing privacy engineering practices (i.e., theories, methods, tools and techniques).

Some of the main limitations found in the related work are already indicated in Table \ref{tab:related-work}.
Among the main research gaps, we noticed that most of the research subjects in the studies come from North American and European regions.
This limitation is a well-known problem in psychology that asserts that the overreliance on Western, Educated, Industrialized, Rich and Democratic (WEIRD) populations can produce false claims about human psychology and behaviour because their psychological tendencies are highly unusual compared to the global population -- and most people are not WEIRD \cite{henrich2010most}.
In this regard, some demographics in studies \cite{hadar2018privacy} and \cite{arizon2021understanding} were also unclear, so we contacted the authors. 
They confirmed that in both studies, the interviewees were mainly from Israel, many of whom work in international (either North American or European) corporations -- so, still not beyond WEIRD.
It is also worth mentioning that six out of the ten studies in the related work (see Table \ref{tab:related-work}) are country-specific, i.e., in which all the interviewed participants live in the same country.
The importance of including more diverse populations and independently replicating prominent studies has already been stressed in the literature, such as the scoping review on OPCC \cite{iwaya2022organisational}.
For such reasons, our research contributes to the existing body of evidence, corroborating and introducing new findings.

In addition, many studies relied on interviews with smaller samples of specific categories of software-related practitioners.
For example, studies focused on senior engineers \cite{bednar2019engineering}, privacy champions \cite{tahaei2021privacy}, and app developers \cite{balebako2014privacy, li2018coconut}.
The work of \cite{hadar2018privacy} also targeted software architects, i.e., directly responsible for the system's design.
However, researchers ought to be careful when defining such inclusion criteria.
Although having power and influence over the actual design and architecture is important, architects usually work with a relatively high-level view of a system.
Developers still need to operationalise things in concrete code and configurations.
We argue that privacy-by-design should adhere to the broadest notion of the word ``Design'', accounting for the entire system design process of analysis, specification, modelling, implementation, testing, deployment, and evaluation of systems.
Any architecture still needs to be translated into low-level engineering.
Governance and data management policies still need to be implemented among the working teams.
Thus, we see the need to interview software practitioners working on all levels of the organisations involved in one or more steps of the software development processes.

Besides, we also noticed that the adoption of privacy engineering strategies is under-reported in the current studies.
Understanding the overall rationale of software practitioners is essential, but we perceived a lack of understanding of the industry's more concrete technical and organisational practices (or lack thereof). 
Hence, our interviews were also intended to investigate specific aspects of software practitioners' knowledge related to state-of-the-art in privacy engineering.
This need goes in line with the emergence of privacy engineering as a field \cite{gurses2016privacy}, and in connection with new standards in the area, such as the ones from NIST \cite{NIST8062, NISTPF}, and ISO \cite{ISO27550}.
So far, many developers have reported that privacy is vague and hard to operationalise, so it is crucial to verify whether they are familiar with and use the current myriad of privacy engineering theories, methods and tools.
If they do not use the existing privacy engineering solutions, it is also crucial to understand why this happens, e.g., lack of knowledge or impracticability.

\section{Methodology}
\label{sec:methodology}

\subsection{Study Design}
This study aims to address the identified gaps in further understanding the software practitioners' perspectives and organisational and engineering practices for dealing with privacy in working systems.
The main Research Question (RQ) that guided this study was: 
\begin{itemize}
    \item \textbf{RQ1}: What are the perspectives on privacy from software practitioners working with systems that process personal data?
\end{itemize}
In addition, this study sought to emphasise privacy engineering by addressing two other sub-questions:
\begin{itemize}
    \item \textbf{RQ1.1}: What is the software practitioners' knowledge of existing privacy laws, frameworks, and engineering approaches?
    \item \textbf{RQ1.2}: What are the organisational aspects reported by practitioners in relation to handling privacy in their work?
    \item \textbf{RQ1.3}: What are the concrete privacy practices used by software practitioners?
\end{itemize}
This emphasis is particularly important now that several privacy laws have been enacted globally, and many advances in privacy engineering are reaching the industry.

To address the research questions, we designed an interview-based study, allowing us to engage and have in-depth conversations with software practitioners on the topic. 
Five instruments were created in the conduction of this study, including: 
(1) a form for registration of interest in the study; 
(2) a screening survey to filter only relevant participants; 
(3) an interview guide covering multiple topics; 
(4) a recruitment strategy, with standard emails and posts; and, 
(5) a complete ethics application (with the participant information sheet, consent form, and data complaints form).

Registering interest in the research was straightforward, allowing participants to read the project description and provide their names and email addresses to researchers.
For the screening survey, the questions were defined to provide a good description of the participant's background, experience in the field, organisational characteristics, and a few points on privacy knowledge.
This survey was designed for two reasons: (1) to filter participants that are software practitioners who work with systems that process personal data, and (2) to capture \textbf{the} demographics from the participants.
The works of \cite{shan2019survey}, \cite{van2019data}, and \cite{bednar2019engineering} were consulted, some questions were adapted, and some were included to create this research's screening survey.
The final screening survey is found in a separate file ``Appendix A -- Screening Survey'', as supplementary material.

The interview guide proposed by \cite{hadar2018privacy} was adapted for this study.
Some questions were included to significantly emphasise the topics on existing privacy laws and privacy engineering theories, methods, and tools.
The modified interview guide is found in a separate file ``Appendix B -- Interview Guide'', as supplementary material.
Standard emails and posts were written to help authors disseminate a short invitation to the study, with a link to the registration of interest form.
When completing the study, participants received a 15 AUD Amazon voucher.
The study proposal and all instruments were included in the ethics application, approved by the University of Adelaide Human Research Ethics Committee (No. H-2020-139).

\subsection{Instruments Review}
Two external researchers were invited to review all the created instruments during the study design process.
These researchers are knowledgeable in qualitative research, especially human factors in cybersecurity and socio-technical systems.
The authors examined all the suggestions and incorporated them by slightly changing questions in the survey and interview guide, enhancing clarity and consistency.
After that, we ran a pilot interview with the updated interview guide.
Based on this pilot interview, we also chose to move the questions about ``Information Sources'' by placing them together with the questions about ``Cases and Examples'', which improved the flow of the interviews.
Our additional questions were also found to be well-positioned within the interview guide.
The interview guide was found to be solid, helping to guide the interview process and elicit further questions along with the interviewees' responses.
With that, all authors were satisfied with all instruments, allowing the study to proceed.

\subsection{Data Collection}
The data collection phase started in August 2020 and ended in January 2021.
Participants were invited through different channels, leveraging our research centres' professional contact networks.
The authors disseminated the project description with the link to the ``Register Your Interest'' form to several contacts, distributing it mainly through mailing lists and social media profiles (i.e., LinkedIn and Twitter).
We purposely did not send bulk emails or search for contacts in external networks (e.g., GitHub) to avoid spamming software practitioners.
Although we do not know the exact number, we estimate that more than a thousand practitioners were invited through this process.
We also know many practitioners extended the study’s invitation to their networks.
In total, 134 had registered their interest, agreeing to receive complete information about the project and a link to the ``Screening Survey''.

Participants were asked to submit their consent forms after answering the screening surveys before scheduling the interview.
A total of 40 participants answered the screening survey.
One participant was excluded because they did not have experience with systems that handle personal data.
The other nine participants dropped out of the study without scheduling their interviews, so their answers to the screening survey were also excluded.
In the end, 30 participants from 29 different companies completed the interview phase.
Interviews were scheduled and conducted via Zoom, lasting 53 minutes on average.
Even though an Amazon voucher was offered to these participants, only 11 participants accepted it.
After that, two researchers transcribed all the interviews while also de-identifying information that could reveal the participants' or their organisations' identities.

\subsection{Data Analysis}
The first author, with the support of another research fellow, was responsible for conducting and transcribing all interviews.
They also spent time familiarising themselves with the data, re-watching interviews, reading transcripts several times and taking notes.
As our main methodology, we used Braun \& Clarke's thematic analysis \cite{braun2006using}, following a primarily inductive approach (i.e., data-driven) to provide a rich and detailed yet complex account of data.
First, we randomly selected three transcripts to start the analysis.
The first author and another research fellow qualitatively coded these three transcripts independently.
The two initial codebooks were then carefully compared by the two researchers.
Equivalent codes were merged, and different codes were discussed and kept in the merged codebook.
This process allowed us to start from a somewhat nuanced number of codes (approx. 142).

The first author then coded the remaining 27 transcripts.
The creation of new codes and sub-codes was constantly documented and discussed.
Weekly meetings were set among the researchers to review the codes generated and, to some extent, to discuss potential themes from the analysis.
Theoretical saturation was observed after coding 22 interviews when little or no new codes were created, i.e., only slightly more nuanced sub-codes.
After all the transcripts were coded, the first author defined the themes and sub-themes based on the final set of codes and ongoing discussions.
The final thematic framework was refined among researchers until an agreement was reached.
Based on that, a complete description of all themes and sub-themes was written.
One reason for not calculating an inter-rater reliability score was that researchers perceived the codes created as only the process, not the product of this thematic analysis, as discussed in \cite{mcdonald2019reliability}.

The researchers also privileged a rich and nuanced account of the data, instead of focusing on the homogeneity of the practitioners' views.
That is, unique statements were captured even if only mentioned by one or two participants, e.g., when identifying privacy practices that are rarely adopted, singular standpoints on privacy and ethics, etc.
Thus, here we also favour the researchers' expertise and nuanced views on the topic instead of pursuing homogeneity via coding agreement.

During the data analysis, the team also found it helpful to employ the Knowledge-Attitude-Behaviour model from social psychology to interpret the practitioner's personal aspects.
Many researchers have already leveraged the KAB model to explain information security awareness, e.g., \cite{thomson1998information, kruger2006prototype, parsons2014determining}.
In brief, knowledge refers to all information that a person possesses or accrues in a particular field of study \cite{schrader2004knowledge}.
Attitude can be defined in terms of three components \cite{schrader2004knowledge}: a cognitive component, such as a belief or idea associated with a psychological object; an affective component of the individual's evaluation and emotion associated with a psychological object; and a conative component, represented as an overt action or predisposition toward action directed toward a psychological object.
Here, the primary psychological object is the participants' construction of privacy.
Lastly, most researchers define behaviour as an observable action \cite{schrader2004knowledge}.

Inspired by the prior work on security awareness based on the KAB model, one of the main themes generated during the data analysis was organised into sub-themes on the practitioners' self-reported knowledge, attitudes and behaviours. 
Such an approach allows a more nuanced description of the theme instead of bundling things in broad terms, such as ``perceptions'' or ``understandings''. 
It also enables a comparison with prior work that points to developers' privacy knowledge (including beliefs and conceptualisations) and attitudes.

\section{Results}
\label{sec:results}
This section provides the study's results and key synthetic findings.
For brevity, we use the notation ($n/30$) along the text to denote a number $n$ out of the $30$ total participants.

\subsection{Demographic Characteristics of Participants}
As shown in Table \ref{tab:demographics-basic}, participants are predominantly male ($22/30$) and also, generally above 30 years old ($22/30$).
Most participants ($21/30$) reside in Sri Lanka, Brazil and Australia, yet other participants are in six other countries.
In addition, Table \ref{tab:demographics-basic} shows that most participants carry out traditional occupations in the software industry, such as engineers, developers, architects, team leaders, etc.
Other occupations ($7/30$) were also represented by individual participants, e.g., data governance manager, business analyst, product director, and application security specialist.
Only one participant reported their professional title as a freelance writer, yet they had also worked as a software engineer and instructor in industry and academia.
The participants were also relatively senior, with 76.7\% ($23/30$) having at least 4 years of experience or more in their fields.
The vast majority also work as full-time employees ($26/30$) for their companies.

\begin{table*}[!ht]
    \centering
    \caption{Participants main demographic characteristics.}
    \label{tab:demographics-basic}
    \begin{tabular}{|llllllr|}
    \hline
        \textbf{Partic. ID} & \textbf{Age Range} & \textbf{Sex} & \textbf{Professional Title} & \textbf{Country of Residence} & \textbf{Employment Type} & \textbf{Experience} \\ \hline
        P01 & 30-34 & Female & Software Engineer & Australia & Full-time & 2-4 years \\ 
        P02 & 25-29 & Male & Application Security Specialist & Australia & Full-time & 4-6 years \\ 
        P03 & 45-49 & Male & Software Engineer & China & Full-time & $>$ 10 years \\ 
        P04 & 20-24 & Female & Software Developer & Australia & Part-time & 2-4 years \\ 
        P05 & 45-49 & Male & Software Developer & Australia & Full-time & $>$ 10 years \\ 
        P06 & 30-34 & Male & Software Developer & Germany & Full-time & 8-10 years \\ 
        P07 & 25-29 & Male & Software Engineer & Sri Lanka & Full-time & 2-4 years \\ 
        P08 & 25-29 & Male & Team Leader & Norway & Full-time & 2-4 years \\ 
        P09 & 40-44 & Male & Team Leader & Australia & Full-time & $>$ 10 years \\ 
        P10 & 35-39 & Female & Freelance Writer & United States & Contract & 6-8 years \\ 
        P11 & 30-34 & Male & UI/UX Engineer & Sri Lanka & Full-time & 4-6 years \\ 
        P12 & 20-24 & Female & Trainee Business Analyst & Sri Lanka & Full-time & 2-4 years \\ 
        P13 & 30-34 & Male & Software Engineer & Germany & Full-time & 8-10 years \\ 
        P14 & 25-29 & Male & Software Engineer & Brazil & Full-time & 8-10 years \\ 
        P15 & 30-34 & Male & Software Engineer & Sri Lanka & Part-time & 4-6 years \\ 
        P16 & 30-34 & Male & Software Engineer & Australia & Full-time & 8-10 years \\ 
        P17 & 35-39 & Female & Data Governance Manager & Brazil & Full-time & $>$ 10 years \\ 
        P18 & 40-44 & Male & Software Developer & United Kingdom & Full-time & $>$ 10 years \\ 
        P19 & 30-34 & Male & Software Engineer & United States & Full-time & 6-8 years \\ 
        P20 & 35-39 & Male & Solutions Architect & Brazil & Full-time & $>$ 10 years \\ 
        P21 & 25-29 & Female & Software Engineer & Sri Lanka & Contract & 4-6 years \\ 
        P22 & 25-29 & Male & Tech Lead & Sri Lanka & Full-time & 2-4 years \\ 
        P23 & 30-34 & Male & Team Leader & Sri Lanka & Full-time & 4-6 years \\ 
        P24 & 30-34 & Male & Software Architect & Brazil & Full-time & $>$ 10 years \\ 
        P25 & 30-34 & Male & Software Engineer & United States & Full-time & 2-4 years \\ 
        P26 & 35-39 & Female & Team Leader & Brazil & Full-time & $>$ 10 years \\ 
        P27 & 30-34 & Male & Software Architect & Brazil & Full-time & $>$ 10 years \\ 
        P28 & 30-34 & Female & Head of Product & Brazil & Full-time & 8-10 years \\ 
        P29 & 30-34 & Male & Team Leader & Sri Lanka & Full-time & 6-8 years \\ 
        P30 & 30-34 & Male & Software Engineer & Singapore & Full-time & 4-6 years \\ \hline
    \end{tabular}
\end{table*}

\begin{figure*}
     \centering
     \begin{subfigure}[b]{0.33\textwidth}
         \centering
         \includegraphics[width=\textwidth]{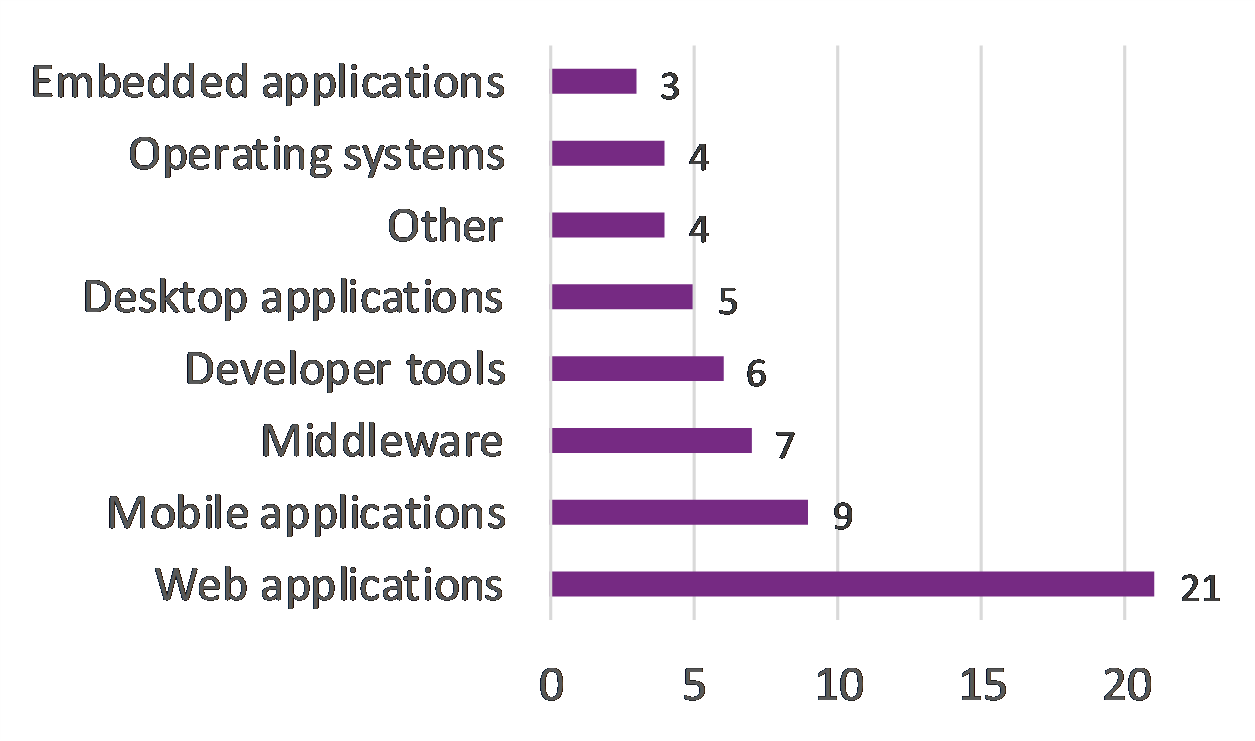}
         \caption{System Types}
         \label{fig:software-developed-prof}
     \end{subfigure}
     \hfill
     \begin{subfigure}[b]{0.35\textwidth}
         \centering
         \includegraphics[width=\textwidth]{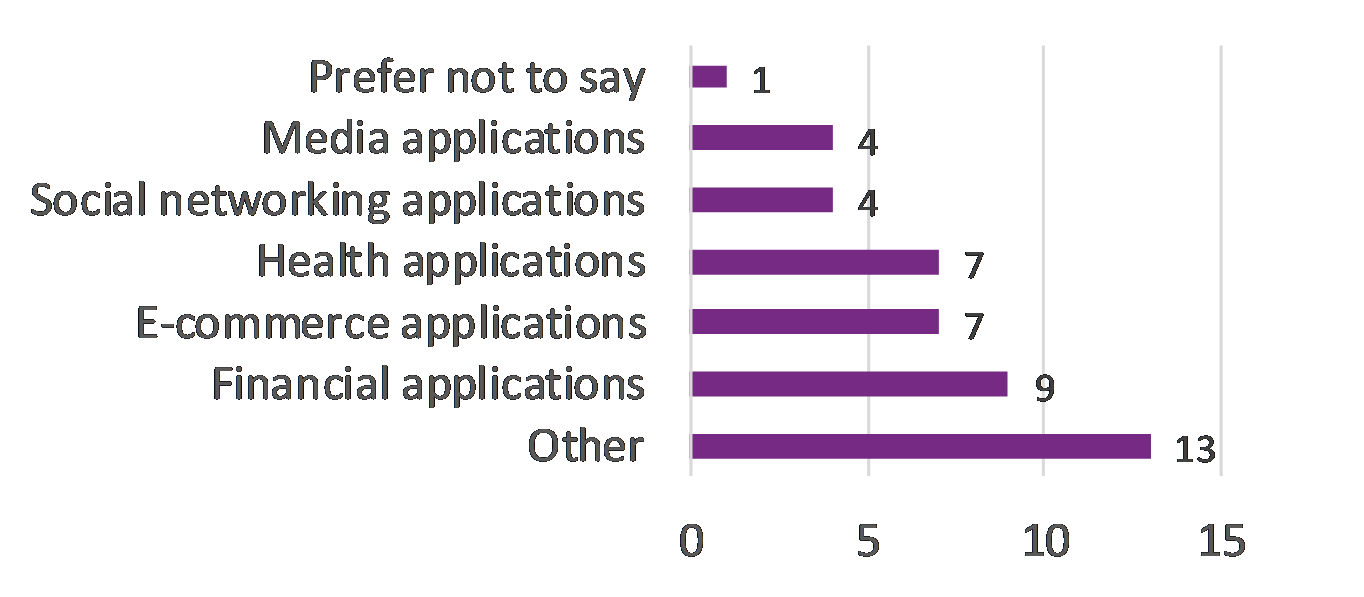}
         \caption{Application Areas}
         \label{fig:types-systems-prof}
     \end{subfigure}
     \hfill
     \begin{subfigure}[b]{0.3\textwidth}
         \centering
         \includegraphics[width=\textwidth]{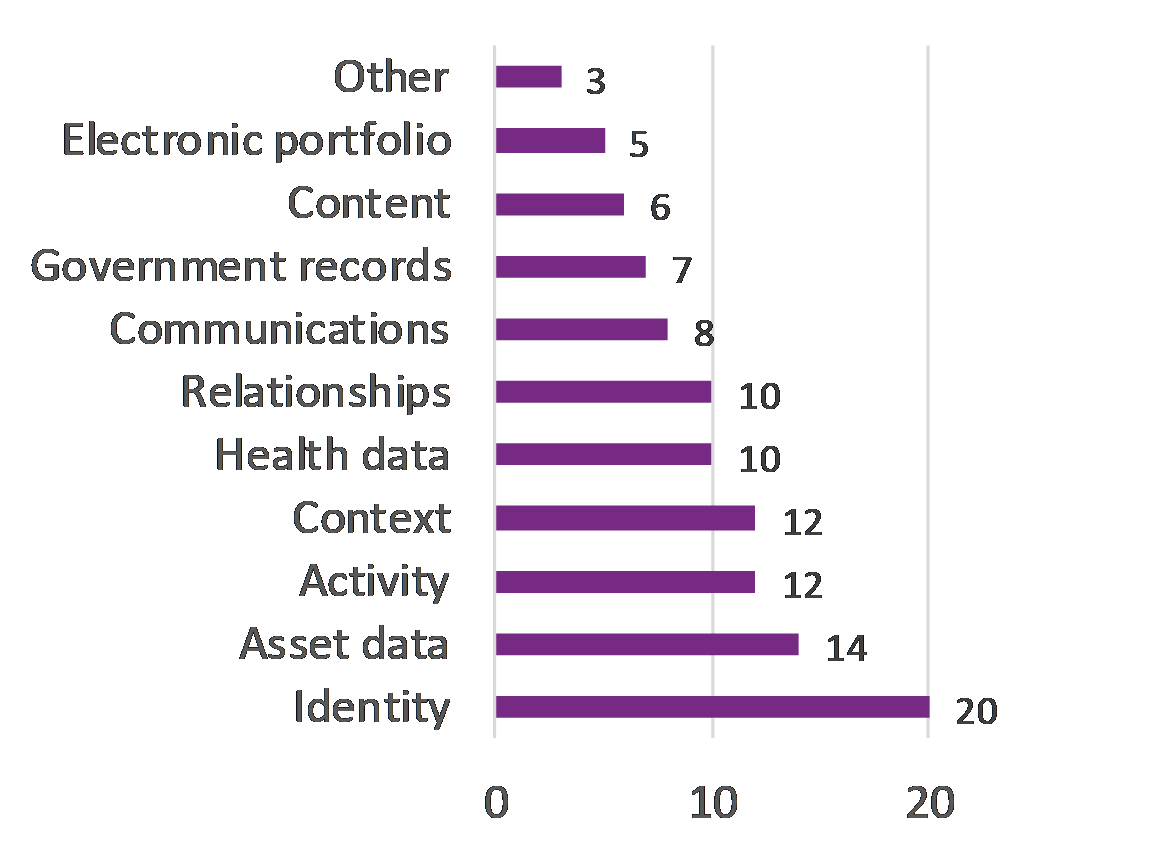}
         \caption{Personal Data Types}
         \label{fig:types-data-prof}
     \end{subfigure}
        \caption{Types of software and applications that handle personal data.}
        \label{fig:types-of-systems}
\end{figure*}

Participants were also asked about the types of systems that they worked with and the application areas.
These were multiple-choice questions, so participants could select as many options as they wanted.
Figure \ref{fig:types-of-systems} shows that the most common type of system was web applications ($21/30$), followed by mobile applications ($9/30$).
Almost all participants reported working with systems that handle either personal data ($27/30$) and/or sensitive personal data ($16/30$).
Only two participants answered that they worked with 'other' types of data.
However, these participants were contacted before scheduling an interview, and it was clarified that they worked with systems that handle personal data.
That is, they only did not have access to any personal data in the development environment.
Participants have worked in many types of applications, such as financial applications ($9/30$), e-commerce ($7/30$), health applications ($7/30$), and many others.
These applications, in turn, processed many types of personal data, such as identity data ($20/30$), asset data ($14/30$), activity data ($12/30$), context data ($12/30$), health data ($10/30$), relationship data ($10/30$), and so on.

\begin{figure*}
     \centering
     \begin{subfigure}[b]{0.3\textwidth}
         \centering
         \includegraphics[width=\textwidth]{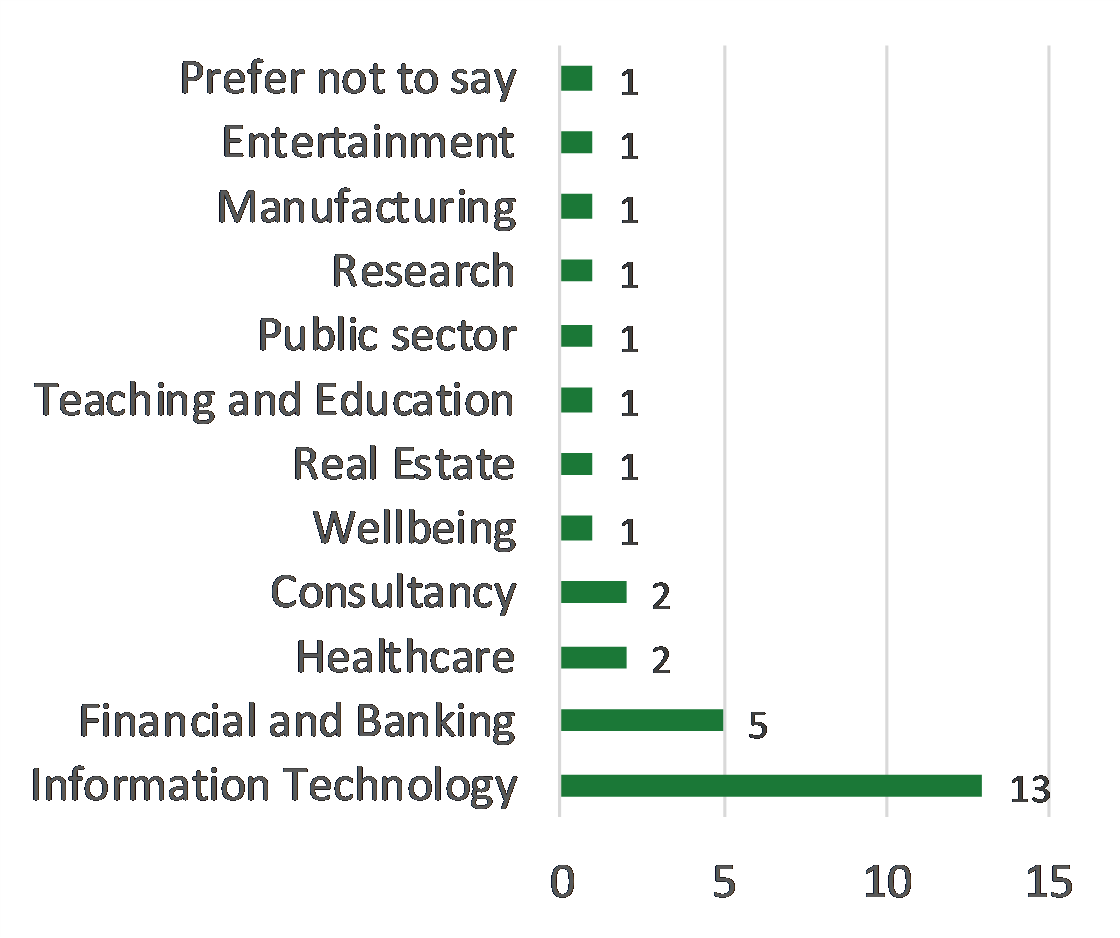}
         \caption{Organisation Domains}
         \label{fig:domain-org}
     \end{subfigure}
     \hfill
     \begin{subfigure}[b]{0.34\textwidth}
         \centering
         \includegraphics[width=\textwidth]{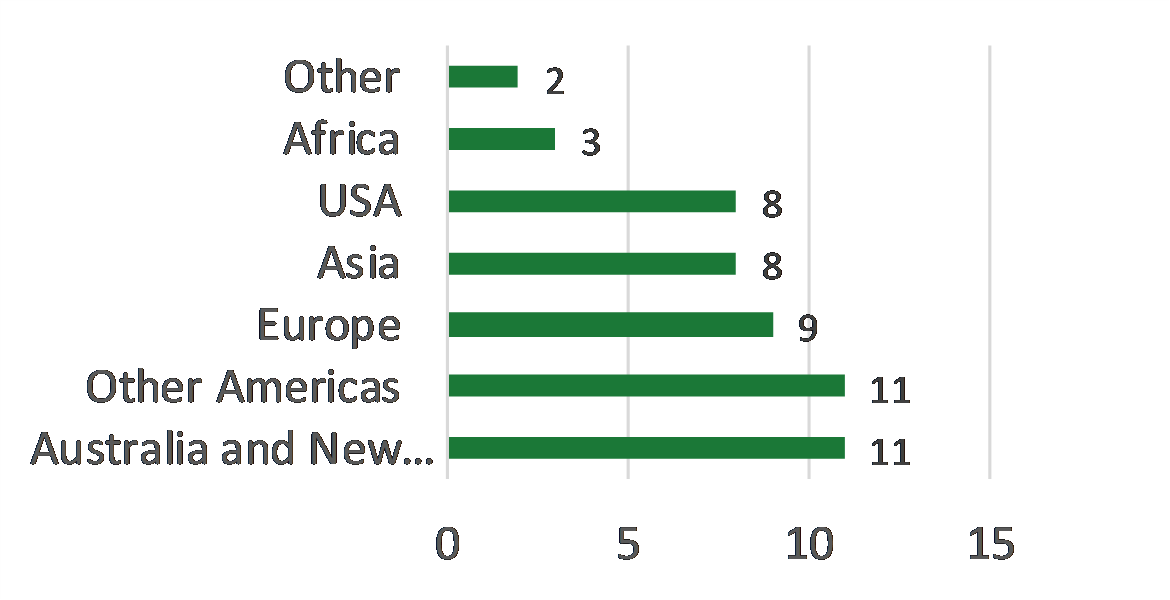}
         \caption{Location of the Clients}
         \label{fig:country-clients}
     \end{subfigure}
     \hfill
     \begin{subfigure}[b]{0.34\textwidth}
         \centering
         \includegraphics[width=\textwidth]{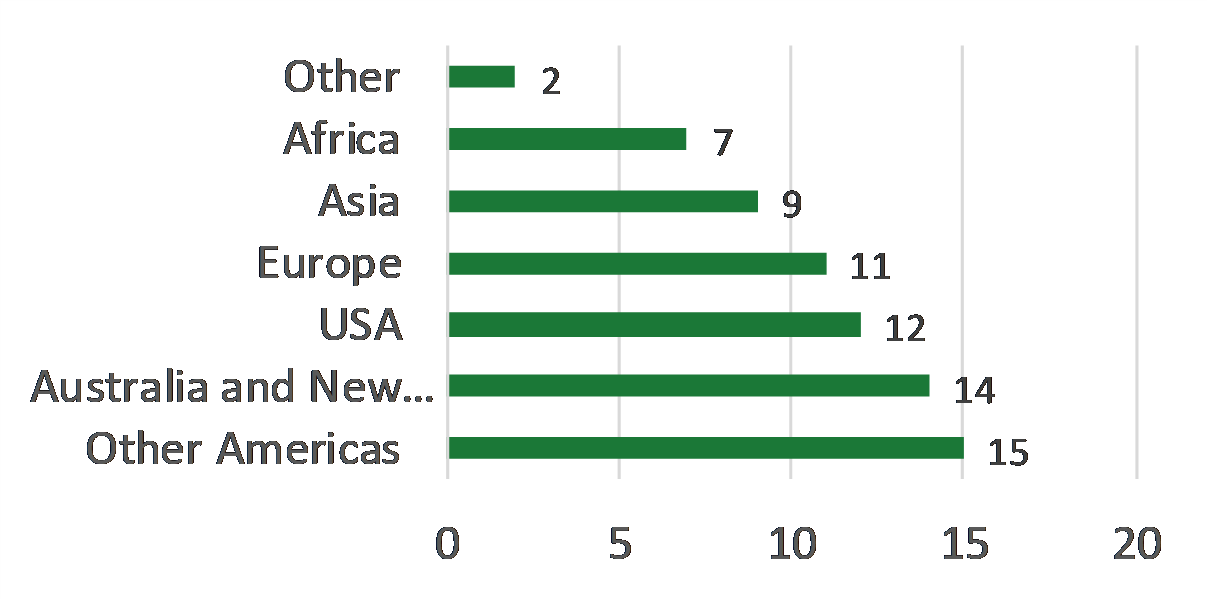}
         \caption{Product Certification Regions}
         \label{fig:country-certifications}
     \end{subfigure}
        \caption{Characteristics of the organisations.}
        \label{fig:types-of-organisations}
\end{figure*}

The participants worked in organisations of different sizes and domains.
Most participants come from large enterprises ($13/30$), followed by solo and start-up practitioners ($11/30$), and SME practitioners ($5/30$) -- only one of the participants preferred not to answer.
As shown in Figure \ref{fig:types-of-organisations}, participants work in companies of many different domains.
Most of them are in the areas of IT ($13/30$) and financial/banking ($5/30$).
These organisations often have international clients and certify/qualify their products worldwide, predominantly in Australia and New Zealand, Europe, the USA and other Americas.

\begin{figure*}
     \centering
     \begin{subfigure}[b]{0.32\textwidth}
         \centering
         \includegraphics[width=\textwidth]{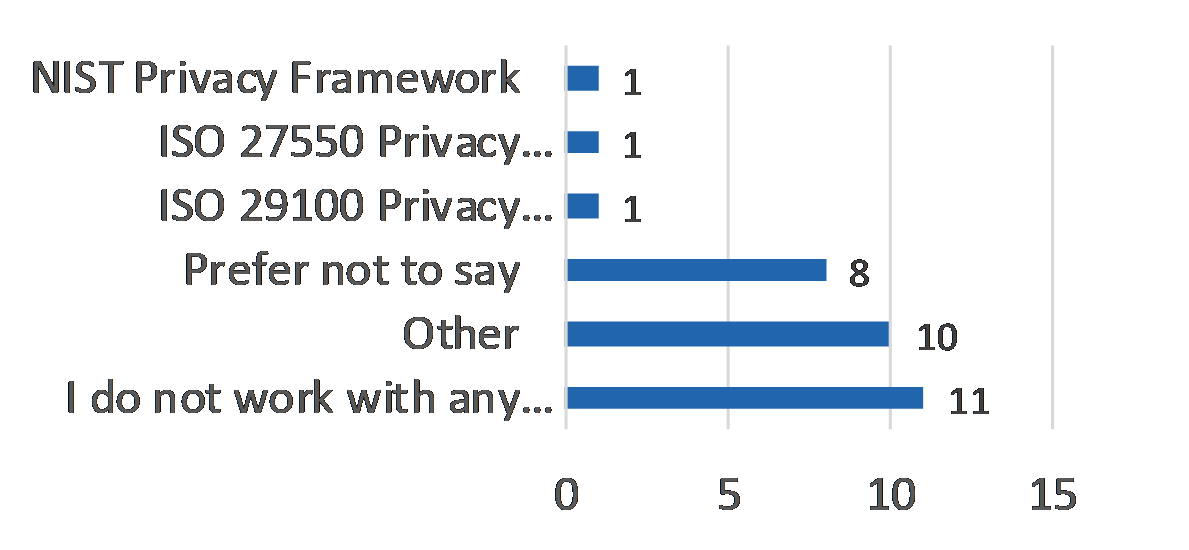}
         \caption{Standards}
         \label{fig:standards}
     \end{subfigure}
     \hfill
     \begin{subfigure}[b]{0.32\textwidth}
         \centering
         \includegraphics[width=\textwidth]{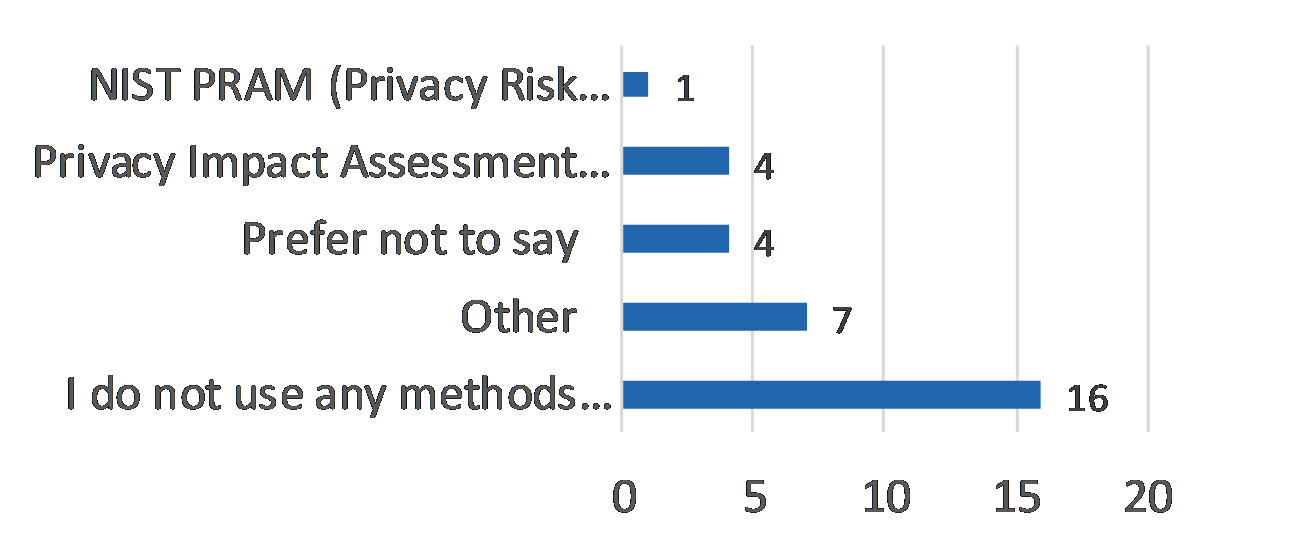}
         \caption{Methodologies}
         \label{fig:methodologies}
     \end{subfigure}
     \hfill
     \begin{subfigure}[b]{0.35\textwidth}
         \centering
         \includegraphics[width=\textwidth]{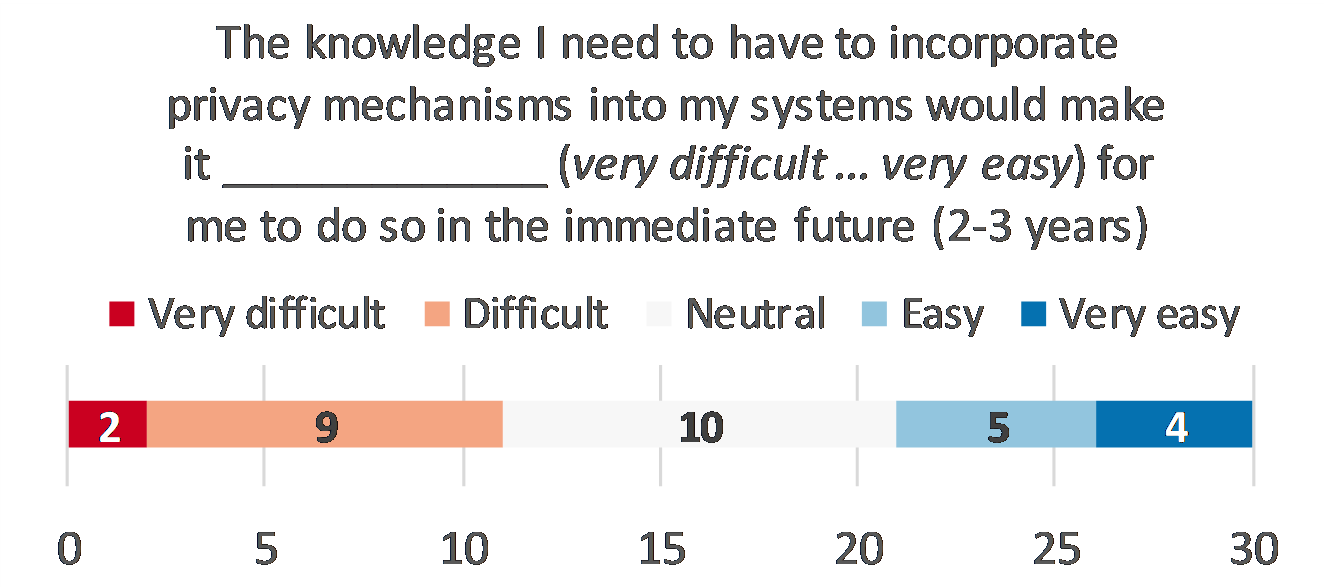}
         \caption{Needed Knowledge}
         \label{fig:needed-knowledge}
     \end{subfigure}
        \caption{Self-reported use of privacy engineering standards and methodologies, and self-assessment on the privacy knowledge to incorporate privacy mechanisms.}
        \label{fig:prieng-answers}
\end{figure*}

Lastly, we also asked participants a few questions regarding existing standards and methodologies for privacy engineering (see Figure \ref{fig:prieng-answers}).
Most participants do not work with any specific privacy-related standard ($11/30$).
Answers under 'Other' ($10/30$) include references to security standards (eg., ISO 27001), privacy laws (e.g., EU GDPR, BR LGPD, AU Privacy Act), as well as \textit{"I'm not sure"}  and \textit{"there are other people working with it"} answers.
However, just two participants specified working with privacy standards, such as the NIST Privacy Framework, the ISO 27550 Privacy Engineering, and the ISO 29100 Privacy Framework.

Similarly, privacy engineering methodologies are not commonly used.
More than half of the participants ($16/30$) stated that they do not use any methods and tools for privacy engineering.
Under 'Other', very few participants individually reported using threat modelling methodologies (e.g., LINDDUN and STRIDE), the AWS Macie data security and data privacy service, information security guidelines, and specific scientific literature (e.g., in anonymisation).
Few participants ($4/30$) reported using Privacy Impact Assessment methodologies.

Participants also self-assessed their knowledge level regarding incorporating privacy mechanisms into the systems they work with within the near future.
Most participants declared that they would find such tasks 'difficult' or 'very difficult' ($11/30$).
A third of them stayed neutral ($10/30$), and the remaining ones found that it would be 'easy' or 'very easy' to do so ($9/30$).

\subsection{Principal Findings}
\begin{figure}[h]
  \centering
  \includegraphics[width=1.0\linewidth]{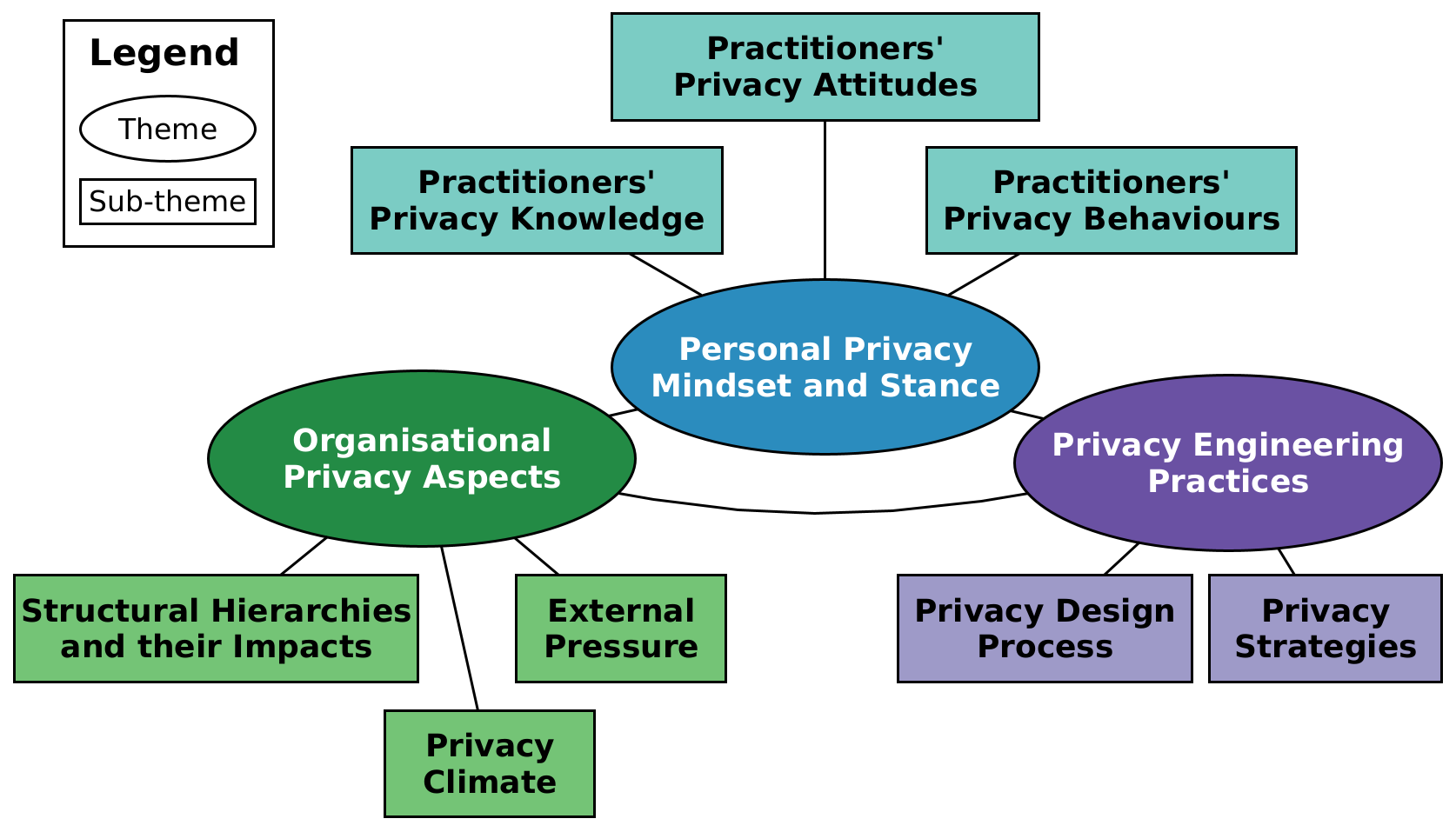}
  \caption{Overview of the study's thematic framework.}
  \label{fig:thematic-framework}
\end{figure}

Here we present the thematic framework that composes the study's narrative through a rich description of the interview data set.
As mentioned, an inductive data-driven approach (i.e., ``let data tell the story'') was followed, allowing the derivation of the themes.
Figure \ref{fig:thematic-framework} provides a model explaining the interrelation among the themes.
It is worth mentioning that the graphical model of the thematic framework is only a simplified and illustrative organisation of the themes and sub-themes that helps to explain the story. 
Themes and sub-themes often overlap and have interrelationships, usually not accounted for in this simplified graphical model.
In brief, the three overarching themes were created in this study:
\begin{itemize}
    \item[(T1)] \textbf{Personal Privacy Mindset and Stance} -- This theme captures the practitioners' personal aspects, organising sub-themes according to knowledge, attitudes and behaviour (''KAB model`` \cite{kruger2006prototype}).
    \item[(T2)] \textbf{Organisational Privacy Aspects} -- This theme captures \textbf{the} aspects of the practitioners' decision power, decision impact, and responsibilities. Also, it covers positive and negative aspects of the ''organisational privacy climate`` \cite{hadar2018privacy}.
    \item[(T3)] \textbf{Privacy Engineering Practices} -- This theme captures how practitioners address and solve privacy in their systems. It also compiles a set of privacy practices and strategies employed by them.
\end{itemize}

Based on Figure \ref{fig:thematic-framework}, a few overarching comments can be made.
First, it is essential to highlight that Organisations are social structures made of people (i.e., practitioners), as viewed by \cite{barrett2006building}.
People can have knowledge, attitudes and behaviours, even if sometimes the knowledge levels and attitudes are not reflected in actual behaviours (e.g., ``do what I say, not what I do'').
In the context of organisational culture, the values and the ``ways of doing things'' are reflections of the people in their social structures, which are especially influenced by the attitudes and behaviours of the leaders.
Considering privacy engineering as just one of the many facets of an organisation, the uptake of privacy values and its engineering theories, methods, tools and techniques will be influenced by the practitioner's and leader's KAB and the ensuing organisation's culture and climate.
As follows, the study's thematic framework is thoroughly described, explaining the themes and sub-themes along with occasional exemplary quotes from the participants.

\subsubsection{Personal Privacy Mindset and Stance (T1)}
\begin{figure}[h]
  \centering
  \includegraphics[width=1.0\linewidth]{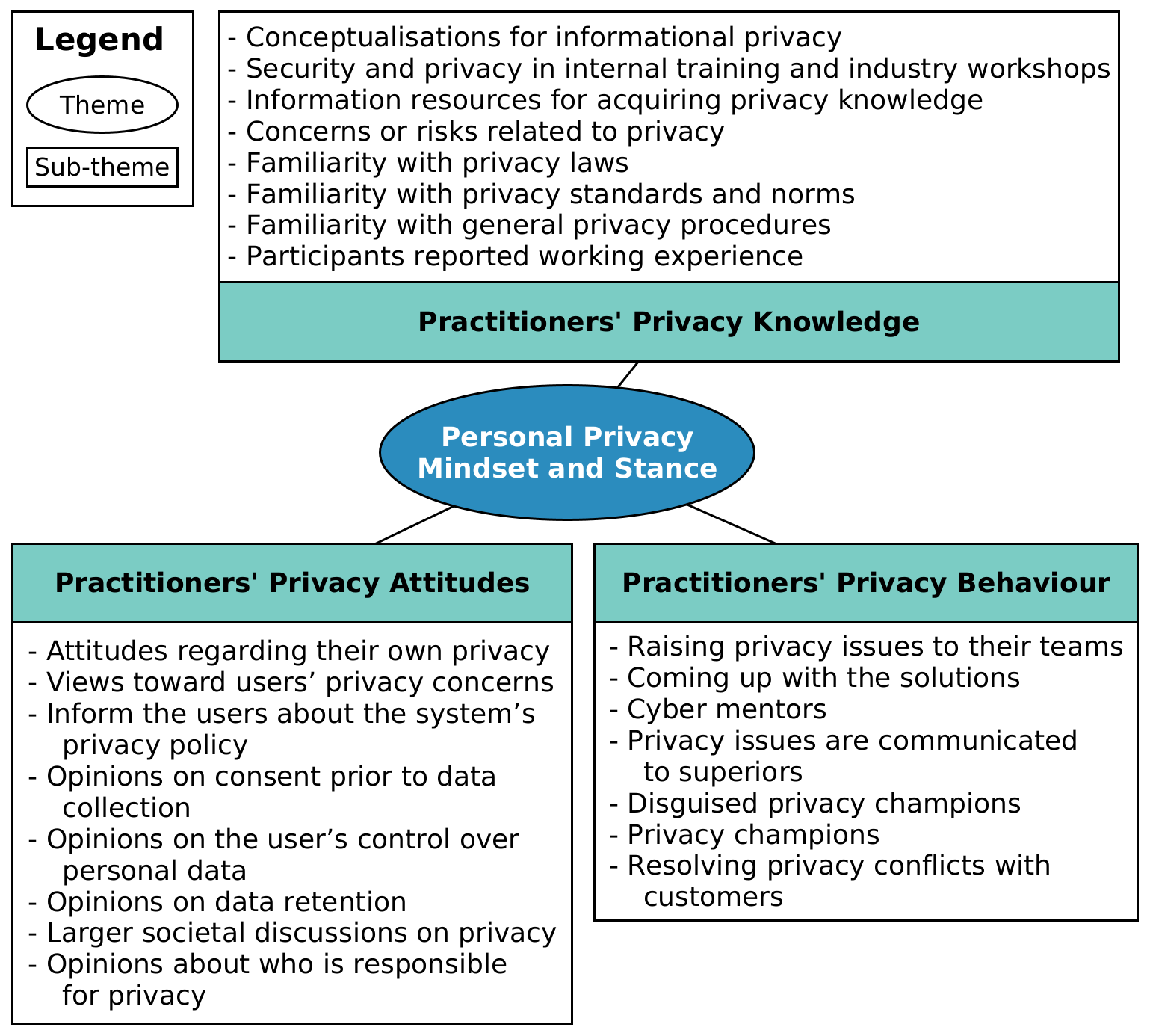}
  \caption{Summary of the theme for Personal Privacy Mindset and Stance.}
  \label{fig:practitioners-kab}
\end{figure}

The most robust theme in the thematic framework reflects the software practitioners' perceptions of informational privacy.
As mentioned, a detailed description of the theme ``Personal Privacy Mindset and Stance'' was structured using the KAB model from social psychology, composed of three dimensions to organise (1) what a person knows (knowledge); (2) how they feel about the topic (attitude); and (2) what they do (behaviour) \cite{kruger2006prototype}.
As shown in Figure \ref{fig:practitioners-kab}, an overview of the theme is provided, depicting a summarised list of topics composing each of the sub-themes.
Most of these topics are also highlighted with bold letters along the text describing the sub-themes.
This figure was drawn as an effort by the authors to provide a more graphical summary of the thematic framework.

\subsubsection*{Sub-theme T1.1: Practitioners' Privacy Knowledge}
Here, evidence of privacy knowledge is captured based on the practitioners' education, training, information resources, conceptualisations of privacy, familiarity with laws and standards, etc.
Most participants were \textbf{highly educated} ($21/30$) with university degrees in computer science, information systems, computer engineering, or other related areas.
Participants also held master degrees ($11/30$) and doctorate degrees ($4/30$) in IT-related fields.
Three participants obtained a degree in other areas, and three participants declared themselves as self-taught practitioners.

Forty per cent of the participants ($12/30$) also reported learning about \textbf{security and privacy in internal training and industry workshops}.
Privacy training in many companies was mandatory, mainly covering the organisations' internal privacy policies, industry-specific regulations (e.g., banking), and privacy regulations (e.g., GDPR).
\begin{quote}
    \textit{``I got some training from my company, such as GDPR. I actually, I was familiar with that, because almost everyone has to take the training. And to take the test, everyone had to pass the test.''} (P03).
\end{quote}

Some participants ($3/30$) were also encouraged by their organisations to join industry workshops and conferences.
For instance, events, quizzes, and awareness programs hosted by the CERTs in their countries.
Very few participants ($5/30$) reported more in-depth training through internal webinars and recorded presentations, e.g., on the usage of data sets for machine learning and privacy design patterns (or ``The Little Blue Book'') \cite{hoepman2014privacy}.
\begin{quote}
    \textit{``We have a [privacy] training program here to train more than 40,000 people in the company [...] It has some different knowledge trails, depending on the person depending on the proximity that to work with personal information [...] And these trails are mandatory for every employee.''} (P17).
\end{quote}

However, most participants ($23/30$) did not report any privacy-specific training as part of their formal education, company internal training, or certifications.
In fact, many participants made a personal effort to learn about privacy so that they would be able to develop systems:
\textit{``The company itself, actually, [...] doesn't have a really structured privacy concern for everybody. So, it's more personal effort.''} (P14).
Some participants ($3/30$) reported extensively reading privacy laws and regulations on their own and consulting with ``policy people'' and privacy experts to see whether they were compliant.

Apart from training and education, participants use other \textbf{information resources} for \textbf{acquiring privacy knowledge}.
Usually, participants ($12/30$) rely on internal documents to inform themselves about privacy.
These include privacy (and security) policies and regulations, general protocols, documentation, internal guidance, and data management procedures.

To a lesser extent, participants ($3/30$) also mentioned reading the relevant regulations, e.g., GDPR and HIPAA, and conducting searches online whenever needed.
Only one participant reported to \textit{``dive back into that literature''} (P02) to solve privacy problems, specifically.
Many turn to their peers as a source of privacy knowledge, getting advice from colleagues in the company ($16/30$), attending conferences in open-source software communities ($2/30$), and following specialists in social media (e.g., LinkedIn and Twitter) ($4/30$).
\begin{quote}
    \textit{``just following the right people and security specialists from around the world, even on social media and stuff like that, is a very good way to learn [...], otherwise we wouldn't get to see [...] within our small little bubbles in our organisations''} (P16).
\end{quote}

Practitioners were also asked about their \textbf{conceptualisations for informational privacy}, and the difference between security and privacy.
Such questions allow us to determine participants' technical knowledge on the topic.
Although most participants ($24/30$) provided rather simplistic privacy definitions, six participants developed more elaborated answers.
That is, a concise answer addressing three or more privacy components (e.g., control over data, informed consent, transparency, data protection, lawfulness, and purpose specification).
In fairness, it is unlikely that practitioners would provide a full-fledged privacy definition or mention fundamental works on privacy, such as Westin's (1968) privacy definition \cite{westin1968privacy} or Solove's (2008) privacy taxonomy \cite{solove2008understanding}.

Additionally, eighteen participants framed privacy as data protection, referring to keeping the data secure, private, undisclosed, and accessed only by authorised parties.
Nine participants emphasised the aspects of privacy as control over personal data, e.g., deciding what is collected, for which purposes, and who has access to it.
To a smaller extent, six participants also expressed privacy in terms of transparency and consent in their definitions.

Regarding the difference between security and privacy, most participants ($14/30$) stressed the focus of security in protecting/safeguarding personal data, e.g., preventing attacks, unauthorised access and data leakages.
On the other hand, a few participants ($8/30$) emphasised that privacy is more than data protection, stressing other dimensions, e.g., making choices, having control over data, providing consent, and respecting ethics.

It is reasonable to say that most practitioners still have a limited and imprecise notion of informational privacy.
However, when practitioners were later asked about privacy strategies (described in Section \ref{sec:privacy-practices}), their answers went beyond such limited definitions, indicating multi-dimensional privacy practices.
So, even though practitioners may not accurately define privacy, they still implement several technical and organisational privacy controls.

Participants also described their \textbf{concerns or risks related to privacy}.
The top concerns for practitioners ($17/30$) were personal data leakages, such as exposing the user's identity, sensitive data, passwords, and credit card numbers due to attacks or accidental disclosure.
Similar risks associated with unauthorised access to personal data, particularly by people inside an organisation, were also mentioned ($7/30$).
Some participants ($7/30$) were also concerned about harming users due to such data breaches, e.g., compromising their safety, creating embarrassment, or discrimination.
\begin{quote}
    \textit{``What can potentially happen to the end-user is that they... It could compromise their safety or otherwise hurt them. Depending on which information got leaked, like you don't want the world to know your address, for example, with your phone.''} (P10)
\end{quote}

Several other privacy risks were expressed by the participants, such as (a) being unable to delete personal data after the users request it ($5/30$); (b) users getting tracked by advertisers ($5/30$); (c) the lack of control when sharing data with third parties ($5/30$); (d) violating the user's consent ($4/30$); (e) violating regulations ($3/30$); (f) mistakenly logging personal data ($3/30$); (g) violating employees' privacy ($3/30$); and, (h) collecting too much data ($3/30$).
Again, participants were not encouraged to enumerate privacy risks exhaustively but to describe their top concerns.

Following up on some of the survey questions, participants talked about their \textbf{familiarity with privacy laws}, standards, and procedures, emphasising the ones they actually used in their organisations.
In this order, the most mentioned privacy laws were the EU GDPR, BR LGPD, AU Privacy Act, and the Health Insurance Portability and Accountability Act (HIPAA).
Although most participants ($19/30$) declared familiarity, they have just superficial knowledge about the laws.
\begin{quote}
    \textit{``I have a vague understanding of that, but I would describe it's vague. I'm aware of the California one. And Brazilian one.''} (P18). \\
    \textit{``So I don't have to worry too much about studying Laws myself. I wouldn't be too worried about that.''} (P19).
\end{quote}

Just seven participants demonstrated a more profound understanding of the privacy laws (e.g., GDPR, LGPD, HIPAA, Australian Privacy Act) by comparing legal frameworks, enumerating privacy principles from the regulations, or discussing the compliance process.
Other regulations, such as the California Consumer Privacy Act and US Privacy Act, were mentioned once.
Besides, four participants considered following the EU GDPR as a best practice, considering it a stricter law that superseded anything before that.
\begin{quote}
    \textit{I think APP [AU Privacy Principles] is kind of soft and rubbery, and doesn't really pinpoint anything. So [...] we like to use GDPR as our core, but then also, [...] elevate a couple of levels to match APP as well.} (P09). \\
\end{quote}

Regarding \textbf{privacy standards and norms}, twenty-four participants either said they were not familiar with standards or just vaguely mentioned following ISO and NIST.
Exceptionally, some participants ($5/30$) specified using standards, such as the Payment Card Industry Data Security Standard, ISO/IEC 27001 for Information Security, Health Level 7, 
South Australia Cyber Security Framework and Protective Security Framework, US Federal Information Processing Standard, and the Common Criteria for Information Technology Security Evaluation.
Interestingly, a few participants ($3/30$) confused privacy regulations as standards, e.g., referring to GDPR and HIPAA, suggesting a lack of knowledge on such aspects.
A mismatch between the survey and the interview answers was also found since nobody mentioned using the NIST Privacy Framework, ISO 27550 or ISO 29100 in the interviews.

Furthermore, instead of talking about specific privacy laws or standards, the participants were also asked about the \textbf{general privacy procedures} they were familiar with.
Seven participants were unaware of any specific privacy procedures used in their work environment.
Here, participants ($11/30$) usually referred to security practices, e.g., access control of databases and encryption/hashing of sensitive information.
The most mentioned privacy practices were data anonymisation, obfuscation and masquerading of personal identifiers (e.g. anonymous IDs, anonymous business metrics), and enabling users to access, request and delete their data from the systems.
This shows again that, in general terms, practitioners adopt a more ``privacy as data protection'' approach based on classic security controls.
Nonetheless, many privacy strategies and procedures were discussed throughout the interviews and are further detailed in Section \ref{sec:privacy-practices}.

The vast majority ($28/30$) of the \textbf{participants reported working experience} with medium- to large-scale systems that process large amounts of data of thousands or millions of users.
Only four participants worked with small-scale software systems, such as personal websites, customer management, and simple game apps.
Most prominently, participants worked in banking and finances (e.g., fintech), clinical and medical systems, e-learning, e-commerce, government services, telecommunications etc.
Also, only six participants had security or privacy as part of their primary tasks, such as making architectural decisions in their projects (e.g., cybersecurity specialist) or building specific solutions (e.g., identity management and personal data inventories).
The other participants would nonetheless participate or be affected by security and privacy considerations, having to deal with it to some extent in their work activities.
\\

\noindent\fbox{
    \begin{minipage}{.95\linewidth}
        \setlength\parskip{1em}
        \textit{\textbf{Synthetic findings:}} Participants reported a lack of formal training on privacy. However, they appear to be more privacy-aware compared to the lack of knowledge and negative experiences reported in previous studies \cite{hadar2018privacy, bednar2019engineering}. Participants are mostly unaware of existing standards related to privacy engineering. Privacy as confidentiality, i.e., emphasis on hiding data and security properties \cite{gurses2010multilateral}, is still the most common conceptualisation.
    \end{minipage}
}

\subsubsection*{Sub-theme T1.2: Practitioners' Privacy Attitudes}
During the interviews, the individuals' self-reported attitudes were captured when they expressed their beliefs and opinions as to rationalise their mindset and stance on privacy topics.
As a result, these attitudes are rather personal, not to be confused with the existing policies and practices in the organisation.
There are misalignments between the participants' attitudes (and behaviours) and the actual practices of organisations, further discussed in Section \ref{sec:organisational-privacy}.

Many participants ($12/30$) talked about \textbf{attitudes regarding their privacy} as users.
The most recurring feeling was the worry of being watched, listened to, monitored, or even manipulated when using apps, social media, or browsing the web.
Examples were the apps listening to their conversations and later showing ads of products that they mentioned.
Also, the extent of data collection by social media, \textit{``I believe that Twitter, Facebook, Instagram, know more about me [...] than myself''} (P20), as well as the psychological manipulation used in marketing strategies, \textit{``I should be actually aware or should choose if you are using the data to manipulate my perception of what's around me''} (P24).
Some participants reported that they \textit{``have been hurt [...] or seen people suffer''} (P11) due to privacy violations, making them further aware of negative impacts.
It was also argued about the importance of having people with such backgrounds, who unfortunately faced some hostility, to understand data privacy risks better.
\begin{quote}
    \textit{``So having those, you know, unfortunately, people with [a] background that is somewhat, you know, faced hostility is important to understand what are those risks of that data, and that, you know, generally is why I try to think about things early, but also be in a position where I can react to it and be willing to make those changes.''} (P16).
\end{quote}

Two participants also expressed their frustration that privacy is unachievable in today's digital society.
As stated, \textit{``to have the comfort of the modern life, we just have to give up our privacy''} (P13).
Sometimes opinions were even more extreme, \textit{``[e]ither you're completely off the grid or you don't have it''} (P24).
Essentially, this evidences a sense of total loss of control over their personal data, as well as facing the dilemma of choice between convenience over privacy.

Three participants also discussed their \textbf{views toward users' privacy concerns}.
Although the general impression is that most users are not concerned about privacy, there is still \textit{``a small subset of users who were very, very aware''} (P18), and will exercise data access and deletion rights.
This worry from the public also spreads among practitioners, especially when privacy violations get publicised in the media, negatively impacting the image of the companies and the image of the developers that work there.
\begin{quote}
    \textit{``there's an honour code involved, like doing good code, because if you're in the company, [and] they leak data [...] it's your fault, right? Nobody wants that. It's like a shaming technique, right?''} (P24).
\end{quote}

On the other hand, the vast majority of participants ($25/30$) said that their organisations \textbf{inform the users about the system's privacy policy}.
Only four participants expressed otherwise, saying that there is no privacy policy for the system or that only European projects require one.
One participant was not sure about it.
However, neither practitioners nor users seem to read privacy policies, or \textit{``if you read the big policy, nobody understands [it]''} (P17).

Following up on that, participants were also asked about their personal opinions on four privacy aspects: user's control over data, informed consent, data retention, and the responsibility for privacy.
According to two-thirds ($21/30$) of the participants, the general \textbf{opinion on the user's control over personal data} is that users should have complete control whenever possible.
However, a third of them ($10/30$) stressed that this level of control depends on the type of system.
Here, they discussed exceptional cases, i.e., for banking, healthcare and law enforcement, in which specific regulations prevent users from accessing or modifying their data.
Another problem raised by three participants refers to the users' capacity to understand the impacts -- especially detrimental ones, e.g., over-sharing data -- that full control over one's data may incur.
So, some level of education would be advised before enabling complete control over data.

Likewise, almost all \textbf{opinions on consent prior to data collection} were rather emphatic ($29/30$), expressing that consent is extremely important, critical, mandatory, and required by law.
Just one participant gave a different opinion, saying that \textit{``[they] don't see the need for that \textbf{if} the system is designed to keep the private data secure and only accessible for those authorised to it''} (P27).
However, two practitioners also expressed some frustration about privacy fatigue, e.g., when repeatedly asking for consent ``gets in the way'' of the user, defeating the original objective.
For instance, when consenting to web cookies, users mindlessly click the accept button to continue navigating.
Also, similar to the user's control over data, a few participants ($3/30$) stressed that depending on the type of system, consent is not required, such as for law enforcement and criminal/fraud investigations.
Nonetheless, when questioned about consent being opt-in or opt-out, twenty-six participants declared that they should opt-in.
Only two participants preferred opt-out to decrease the friction in the onboarding process but admitted that the users should still be informed.
The two remaining participants believe that opt-in and opt-out for consent depend on the system.

Participants also gave their \textbf{opinions on data retention}.
Although almost half of the participants ($14/30$) agreed that personal data should definitely be deleted at some point, most of them stressed that it depends on the type of data and system.
Again, data retention might be compulsory for banking and healthcare systems, but they agree that these are exceptional cases.
In general, systems can delete transactional data or data in-use as soon as possible or when they are no longer required.
However, participants could not specify a precise time frame for data retention since it seemed somewhat circumstantial.
About half of the participants ($14/30$) suggested data retention periods between one and three years as a rule of thumb.
Two suggested anonymising personal data instead of deleting it to be still utilised for business purposes.

The \textbf{opinions about who is responsible for privacy} in the organisation were also shared by the participants.
Thirteen of them consider privacy as a collective responsibility for the entire organisation.
Notwithstanding, twelve participants also expressed that the system's architect bears significantly more responsibility for privacy.
Developers, business analysts, project managers, and privacy officers also share this responsibility but were mentioned to a much lesser extent.
Some practitioners ($5/30$) also argue that their customers are responsible for privacy since they ultimately decide on the requirements that get implemented or deployed and operate the systems independently.
Interestingly, eight participants expressed a personal sense of responsibility for developing privacy-aware systems.
These participants also mentioned a sense of duty, identifying and raising privacy concerns that they see lacking in their organisations or taking leadership roles regarding privacy. 

Interestingly, about a third of the participants ($11/30$) initiated many \textbf{larger societal discussions on privacy}.
For instance, they ($4/30$) discussed their impression of a post-privacy world, where privacy is not guaranteed or expected anymore.
This is associated with the aforementioned feeling of privacy being unachievable in today's digital society.
Four participants pointed to the clash between business profit and privacy for most big tech corporations in a capitalist society.
Personal data is treated as an asset that tech companies, advertising platforms, and data brokers can monetise.
On top of that, it is hard for the developers to verify that big-tech service providers are processing the data only for specified purposes, deleting it when requested, and so on.
\begin{quote}
    \textit{``From my point of view, I don't need that data at all. Then the question is, has that data been deleted? [...] there's just no way I have been able to go and ascertain whether it's deleted or not, even if they wanted to, there's no real way of doing that. Because each little packet of information is so interconnected to everything else and I don't see how they can do it easily. And then it is not encouraged to do obviously [...]''} (P18).
\end{quote}

On the other hand, some practitioners ($4/30$) also \textit{``think [that] there's been a change of mindset in this space''} (P28).
They say that they are more consciously thinking about every piece of information they collect, why they need it, and who will access it.
There seems to be a movement for a privacy renascence, accompanied by the new wave of privacy legal frameworks and standardisation efforts, and a shared sense of responsibility from the practitioners.
\begin{quote}
    \textit{``Because we've become so accustomed to just sharing all that data and all that data being collected by, you know, your Googles and the Facebook's of the world, and use the very targeted ads. [...] maybe we sort of bounce back and sort of say, actually, you know, we've sort of let this go on too long. We need to put more regulations in place and educate developers and things like that more to sort of protect privacy.''} (P02)
    \\
    \textit{``Privacy is [...] about a social contract between you as the software owners and the person whose information you're holding and upholding the terms of that social contract.''} (P09)
\end{quote}

\noindent\fbox{
    \begin{minipage}{.95\linewidth}
        \setlength\parskip{1em}
        \textit{\textbf{Synthetic findings:}} While participants value their privacy, mixed feelings exist that ``privacy is dead'' and that most users are unconcerned. Participants generally think that users should be in control over their data and that user consent must be collected in most cases before data collection. Privacy is often considered a shared responsibility with a heavier load carried by software architects.
    \end{minipage}
}

\subsubsection*{Sub-theme T1.3: Practitioners' Privacy Behaviour}
The participants' behaviour is captured with emphasis on their self-reported independent actions concerning privacy in their work activities.
However, the practitioners' privacy behaviour significantly overlaps with organisational and privacy engineering practices.
Therefore, to avoid redundancy, this section focuses on the privacy behaviours unique to the participants (i.e., personal behaviours). 
Nevertheless, the theme of privacy engineering practices can be considered a superset of this sub-theme.

The most common behaviour from participants was \textbf{raising privacy issues to their teams} when faced with such problems.
Most participants ($24/30$) reported that they had initiated discussions on privacy during professional activities.
Only six participants stated that they had not been in a position where they felt the need to raise privacy concerns.
Usually, the raised \textbf{privacy issues are communicated to superiors} ($25/30$) by bringing them up to the project leaders, more experienced colleagues or alerting security operation teams.

\begin{quote}
    \textit{``Since I was given this requirement from the person in charge of marketing. One of the things that I did was we go through the privacy policy because I've come across an advertising ID collection idea for a collection, and it's a privacy issue.''} (P08).
    \\
    \textit{``When I found the [privacy] issue, I talked to the team leader, which understood the problem and planned the fix with the team. Later we talked about it in our company chat app so other teams could learn from that privacy concern.''} (P27).
\end{quote}

Besides raising privacy issues, practitioners are also \textbf{coming up with solutions} themselves.
Seventeen participants described moments when they faced privacy problems and how they went about solving them.
For instance, when they identify insecure practices (e.g., in code reviews) and propose safer design strategies, protocols, or storage solutions.
They also write up and change the system's privacy policy when incongruities are found.
Or when dealing with privacy in the requirements engineering stage, communicating with clients, and even setting up secure and privacy-aware architectures or data governance models for the entire organisation.
However, even though many practitioners may be able to raise privacy issues, problem-solving often depends on the organisations' superiors or dedicated security and privacy teams.

One way for managers to offload privacy decisions is by raising privacy awareness and empowering practitioners in the organisation.
In this regard, it was found that some people in their organisations behave as \textbf{disguised privacy champions}\footnote{This concept of a \textbf{privacy champion} refers to key people in an organisation (not necessarily privacy experts) who advocate, encourage others, facilitate conversations, build a culture and help to solve privacy problems. Privacy champions can be compared to other roles, e.g., ``security champions'' and ``innovation champions'', as further discussed in the work of \cite{tahaei2021privacy}.}.
A total of eight participants reported being or having colleagues who are common contact points in the team to get advice and help solve privacy issues.
Besides that, three participants assumed official roles of \textbf{privacy champions} in their teams or companies.
Two other participants were also designated as \textbf{cyber mentors} for junior developers, advising primarily on security but also privacy to a lesser extent.
Practitioners acting in such roles are essential for raising privacy awareness among team workers and translating privacy principles into concrete engineering tasks.

Lastly, some participants ($4/30$) also expressed behaviours when \textbf{resolving privacy conflicts with customers}.
In two cases, practitioners had to persuade the customer either to implement privacy controls in the system or oppose function creeps that would violate privacy.
Naturally, customers may see privacy controls as extra cost in their projects, but the recent privacy laws facilitate these conversations since organisations may be obliged to comply.
Exceptionally, one participant reported facing suspicious requests for excessive data collection, of which the customer would not clarify the purposes, so they had to refuse the implementation.
\\

\noindent\fbox{
    \begin{minipage}{.95\linewidth}
        \setlength\parskip{1em}
        \textit{\textbf{Synthetic findings:}} The most commonly reported behaviour is raising privacy issues with the team and superiors. Roles such as privacy champions are beneficial to software teams, helping others with advice and solutions for privacy issues.
    \end{minipage}
}

\subsubsection{Organisational Privacy Aspects (T2)}
\label{sec:organisational-privacy}
\begin{figure}[h]
  \centering
  \includegraphics[width=1.0\linewidth]{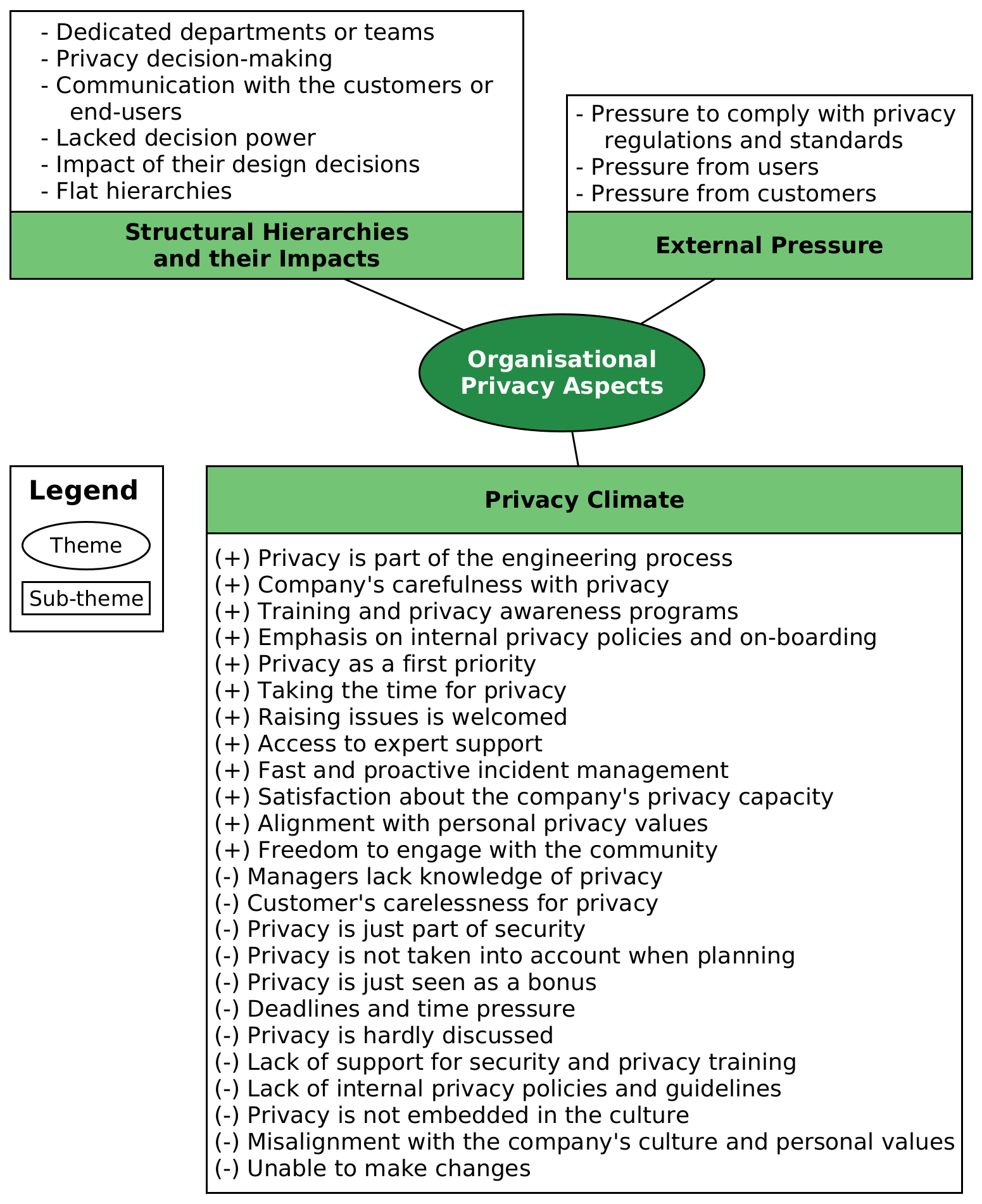}
  \caption{Summary of the theme for Organisational Privacy Aspects. Note: the Privacy Climate sub-theme is composed of positive (+) and negative (-) talk.}
  \label{fig:organisational-privacy}
\end{figure}

Organisational factors, such as culture and climate, influence the practitioners' behaviour and engineering practices.
This theme, therefore, captures some of the components of organisational privacy aspects as reported by the participants.
It is worth reminding, however, that the study's participants are typically bound to confidentiality agreements that prevent them from disclosing organisational factors.
Nonetheless, this study's interviews enabled many components regarding the organisations' culture, climate, and external pressures to be captured in a theme and sub-themes.
Figure \ref{fig:organisational-privacy} provides the theme's graphical summary with the main topics that compose the sub-themes.

\subsubsection*{Sub-theme T2.1: Structural Hierarchies and their Impacts}
The vast majority of the participants ($24/30$) work in small agile teams, regardless of the organisational structures and sizes.
As aforementioned, practitioners often consult with superiors, e.g., team leaders and managers, when faced with privacy issues.
Apart from that, about two-thirds of the participants ($19/30$) also said that their organisations have a \textbf{dedicated departments or teams} to handle security and privacy concerns.
These are (a) legal teams, (b) security teams, or (c) privacy or data protection teams.
It is also common to see legal and security teams absorbing a privacy division.
In fact, dedicated privacy teams were only mentioned by eight participants, as a data steering committee\footnote{The term \textbf{data steering committee} was used in the context of ``data governance'', in which a group of key people in the organisation discuss, plan and coordinate the implementation of data governance strategies, which includes data privacy.} or a data protection team\footnote{The term \textbf{data protection team} or privacy team refers to a particular group of people that is responsible and helps others to address privacy problems in the organisation. Such teams comprise data privacy officers, lawyers, privacy experts or consultants.}, but also as a single privacy champion, data privacy officer, or external consultant.

Although dedicated privacy teams exist, most participants also initiate discussions ($24/30$) on privacy, and all of them can play a role in the \textbf{privacy decision-making}.
Among the participants, one-third reported ($10/30$) that their design decisions affect several people in many development teams, from a few dozen to hundreds of developers.
Other participants ($18/30$), mostly in start-up companies, also reported that the \textbf{impact of their design decisions} often affects the entire team.
Six participants explicitly described a welcoming culture for discussing and addressing privacy concerns in their systems.
Only two participants expressed that they \textbf{lacked decision power} to make the changes, even though they were aware of privacy issues.
Some exceptionally progressive cases come from participants in large open-source software companies with rather \textbf{flat hierarchies}.
\begin{quote}
    \textit{``[...] there's no hierarchy or superiority or one person gets to overrule the other. We really try to make sure that in our work that when someone raises a question or a comment or concern, all of these [privacy] things are addressed, especially in code reviews [...]''} (P16).
\end{quote}

Practitioners ($22/30$) also influence privacy decisions during their \textbf{communication with customers or end-users}.
That is, they can discuss privacy during the system's requirement analysis, testing, and deployment.
Fourteen participants developed systems for specific customers, such as customised solutions and business management software (e.g., ERPs).
In such cases, practitioners can take a more active role in privacy discussions since \textit{``the customer [...] doesn't know how to build software''} (P24).
Sometimes clients even have to be convinced about implementing security and privacy controls.
\begin{quote}
    \textit{``when we're going through the requirements gathering process, and I'll tend a few words about security [...] it's never really a long and detailed discussion, and I've never had a customer who wanted to go further''} (P05).
\end{quote}

Only eight participants reported not having direct communication with their customers or end-users.
Therefore, they rely on other business analysts, managers, and customer support channels to understand and clarify privacy requirements.
Examples of exceptionally privacy-aware customers are financial and telecommunication organisations in heavily regulated markets.
In such cases, practitioners ($2/30$) reported receiving extensive support when dealing with security and privacy requirements.
\begin{quote}
    \textit{``[...] they also provide us [with] some checklist [...] we will make sure we match the checklist. And during this process, they will also send you some experts into our project and guide us to follow this regulation''} (P03).
\end{quote}

\noindent\fbox{
    \begin{minipage}{.95\linewidth}
        \setlength\parskip{1em}
        \textit{\textbf{Synthetic findings:}} Most organisations have dedicated teams handling privacy concerns, yet they are usually represented as a legal or security division. Participants generally feel that they have the power to influence decisions and that raising privacy issues is mostly welcomed in their teams.
    \end{minipage}
}

\subsubsection*{Sub-theme T2.2: Privacy Climate}
This sub-theme is articulated based on Hadar's \textit{et al.} \cite{hadar2018privacy} definition of \textit{``organisational privacy climate as a shared perception of the way behaviour with regard to privacy is rewarded, supported and expected.''} \footnote{This definition was inspired in the work of Schneider \textit{et al} \cite{schneider2013organizational}, that previously posed: \textit{``Organisational climate may be defined as the shared perceptions of and the meaning attached to the policies, practices, and procedures employees experience and the behaviours they observe getting rewarded and that are supported and expected.''}}.
During the interviews, participants also described many aspects that positively or negatively affect how privacy is handled in their organisations.
Therefore, such perceptions were captured and categorised as positive or negative talk, presented as follows.

Some elements in the \textbf{positive talk (+)} have already been mentioned but are worth recapitulating here.
In total, eighteen participants expressed positive perceptions regarding their organisational policies, practices and procedures that are encouraged and rewarded.
The idea of having \textbf{privacy as part of the engineering process} ($8/30$), sometimes called privacy-by-design, was one of the most prevalent signs of a positive privacy climate.
That means the companies have clear procedures and a shared view of addressing privacy issues early in projects and processes.
Remarkably, a couple of participants stressed that such value for privacy might even create contention with customers, requiring them to raise privacy awareness and sometimes push for privacy as a non-negotiable in the projects.

Another aspect of the positive privacy climate captured was the perception ($7/30$) that their \textbf{companies are careful and serious about privacy}.
Many participants ($6/30$) emphasised it as a company value, evidenced by visible artefacts such as \textbf{privacy training and awareness programs}, and \textbf{internal privacy policies} ($5/30$) to be followed, especially during the induction of new members.
Some striking examples are practitioners' perceptions ($2/30$) that \textbf{respecting privacy is a first priority} (e.g., privacy is more important than user base growth, 
the company takes the time for privacy).
One practitioner even mentioned employment contract clauses that protect them in case they act against the company's interest but in favour of ethical values, such as the right to privacy.

\textbf{Welcoming participants to raise issues and initiating} discussion around privacy was also positively perceived by the participants ($6/30$).
In addition, treating employees as equals (e.g., flat hierarchies), incentives for collective decision-making and consensus, and embracing opinions from an international and diverse group of people were all perceived as contributing factors to a positive privacy climate.
Together with that, having \textbf{access to experts and mentoring strategies} for privacy was also seen as necessary ($3/30$).

Even though not as prevalent, a few other perceptions expressed by the participants exemplify behaviours that are just as important for a positive privacy climate.
For instance, \textbf{expressions of great satisfaction} ($3/30$) with the company's current state of practices on privacy, and thinking that \textit{``we have [privacy] pretty much covered''} (P19) or \textit{``we are as good as we can be with respect to privacy''} (P06).
In some cases, participants even perceived their organisations to have surpassing levels of privacy expertise, i.e., to the point where even external privacy lawyers and consultants would not know better.
Also, the perception ($3/30$) of having a \textbf{fast and proactive incident management process} in which identified issues are rapidly resolved.
The perception ($3/30$) of \textbf{alignment between the individual's and the company's values} on privacy contributed to the appreciative and proud feeling of being part of that organisation.
Lastly, the \textbf{freedom within the company to engage} with internal and external communities to learn and discuss privacy topics (i.e., such as participating in this study, conferences, or OSS groups) was also perceived ($1/30$) as positive.

On the other hand, evidence of \textbf{negative talk (-)} was also expressed by thirteen participants.
The most prevalent ($3/30$) was a perception that managers and other \textbf{superiors had a lack of knowledge of privacy}.
This creates frustration since practitioners may have to convince leaders to incorporate technical and organisational privacy controls.
The \textbf{customer's carelessness for privacy} was also badly perceived ($1/30$), such as when privacy is seen as an added cost or when customers lack resources for putting privacy controls in place (e.g., not having a Privacy Officer).
Upper management also confuses the concepts and regards \textbf{privacy as just as part of security} ($1/30$), which leads to miscommunication.
\begin{quote}
    \textit{`Privacy is not a separate entity as such, it's always embedded into the security things. So if you ask the [Company], what are your privacy controls? They'd be like, what? They wouldn't know.''} (P01).
\end{quote}

Practitioners ($3/30$) also perceived that \textbf{privacy is not taken into account when planning} for future requirements.
One participant remarked that \textbf{privacy is just seen as a bonus} in the development team, i.e., an afterthought once everything else is working.
Two participants also pointed to \textbf{deadlines and time pressure} as detrimental factors, which are also related to bad planning and poor design when it comes to privacy.
For some, their perception is that \textit{``privacy is hardly discussed in [the] company''} (P27), and that it is not embedded in the culture.
Also, there are no systematic processes for addressing privacy concerns when they arise and no incentives for acquiring additional security and privacy training.
Two participants also pointed to the \textbf{misalignment of their personal values} and the company's organisational privacy culture.
In one instance, a participant reported feeling frustrated with the lack of privacy concerns, being \textbf{unable to make the changes} needed in the systems, resulting in their resignation.
\\

\noindent\fbox{
    \begin{minipage}{.95\linewidth}
        \setlength\parskip{1em}
        \textit{\textbf{Synthetic findings:}} Signs of a positive privacy climate are addressing privacy at the design stage, regarding privacy as a company value, allowing issues to be raised, and a feeling of alignment between individual and company values for privacy. Conversely, a negative privacy climate was shown through the superiors' lack of knowledge about privacy, views of privacy as an extra cost (low value and low priority), and the lack of incentives for privacy training.
    \end{minipage}
}

\subsubsection*{Sub-theme T2.3: External Pressure}
The \textbf{pressure to comply with privacy regulations and standards} was among the primary concerns according to twenty-three of the interviewees.
Privacy laws such as the EU GDPR and the BR LGPD were the most mentioned, followed by the AU Privacy Act and the US HIPPA.
A few participants ($5/30$) named standardisation bodies (e.g., ISO and IETF) and specific standards (e.g., Federal Information Processing Standards, Request for Comments, Payment Card Industry Data Security Standard, and the Common Criteria).

Although privacy laws were often seen as a positive change in the technological landscape, many participants ($9/30$) reported that compliance with multiple regulations and standards has been very challenging.
Compliance with regulations, such as the EU GDPR, is especially onerous for small and medium enterprises that usually lack the resources to cope with the legal complexities and implement privacy controls.
In such cases, practitioners ($7/30$) sometimes perceive the privacy laws as a burden that stifles innovation and makes them less competitive than larger corporations that can handle such heavy regulatory compliance demands.
\begin{quote}
    \textit{``So okay, now stop whatever innovation you're doing, and just comply. We've got all of this and it was a lot of things to handle. It took us like, six months to do everything.''} (P28).
    \\
    \textit{``And then you have to sort of stop and think about how we're going to change the app now to meet these obligations. And that's quite onerous. I would say it's difficult. It's one of those things where it takes a lot of time and energy with no return on investment at all [...]''} (P18).
\end{quote}

To a smaller extent, the \textbf{pressure from customers} and the \textbf{pressure from users} for privacy compliance were also mentioned by participants ($2/30$).
Customers most concerned about privacy usually have subcontractors, either developing the system for them (e.g., a software house) or working as a service provider (e.g., cloud computing).
The customer would have specific demands for privacy compliance, shared responsibility models, and sometimes auditing processes in such situations.
Only three participants commented on the pressure from end-users, usually requesting access or deletion of their data in the systems.
Therefore, according to the data, regulations drive the integration of privacy controls and (to a lesser extent) compliance requirements from customers to their subcontractors and privacy-aware users.
\\

\noindent\fbox{
    \begin{minipage}{.95\linewidth}
        \setlength\parskip{1em}
        \textit{\textbf{Synthetic findings:}} Privacy compliance is a primary concern for most organisations, but it is especially burdensome to smaller companies with limited resources.
    \end{minipage}
}

\subsubsection{Privacy Engineering Practices (T3)}
\label{sec:privacy-practices}
\begin{figure}[h]
  \centering
  \includegraphics[width=0.8\linewidth]{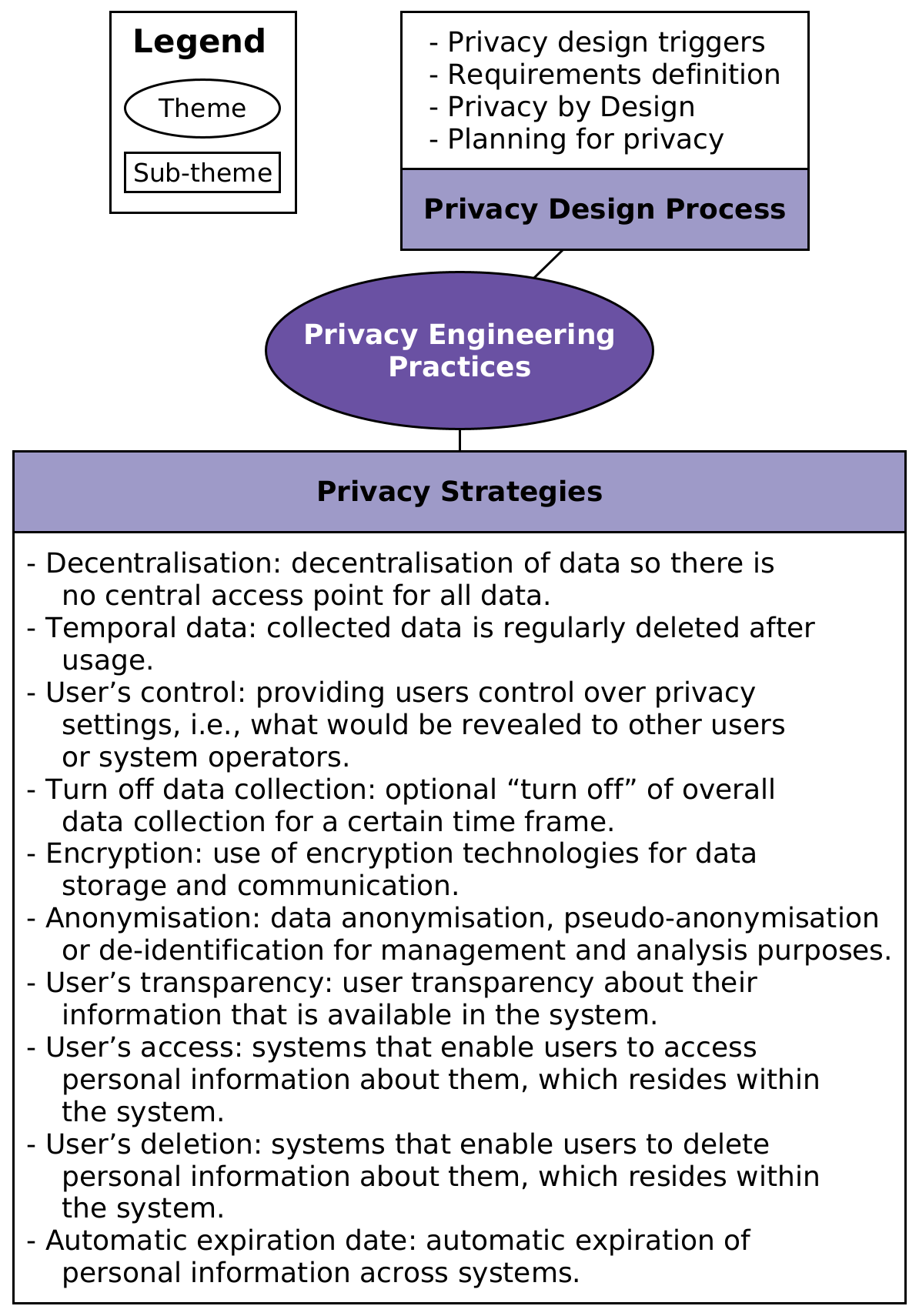}
  \caption{Summary of the theme for Privacy Engineering Practices.}
  \label{fig:priveng-practices}
\end{figure}

The analysis of the interviews' data revealed several practices used by the participants to address privacy concerns in their organisations and in the systems engineering process.
These practices include personal behaviours, organisational procedures, and privacy engineering components (i.e., theories, methods, techniques and tools) used to address privacy issues.
Therefore, two sub-themes were created: 
i) privacy design process that captures the participants' approach for dealing with and solving privacy problems; and, 
ii) privacy strategies, capturing specific management, operational, and technical privacy controls.
Figure \ref{fig:priveng-practices} provides this theme's graphical summary with its main topics.

\subsubsection*{Sub-theme T3.1: Privacy Design Process}
Twenty participants mentioned specific \textbf{privacy design triggers} that made them start thinking in terms of privacy and data protection.
The most common triggers were whenever a new project or idea started ($6/30$), i.e., at the early stages of the design process or when collecting sensitive data.
A few participants ($5/30$) stressed that privacy is part of the feature development process, pertinent during the whole development life cycle.
Whenever a feature is added, changed, or fixed, they are mindful of releasing it without any privacy violations, particularly considering the system's privacy policy.
Interestingly, two participants also mentioned that they consider privacy at the deployment stage, e.g., when releasing a new app that uses 3rd-party cloud services or when exposing RESTful microservices.

As aforementioned, some customers can elicit privacy requirements, but usually, the \textbf{requirements definition} is left to the practitioners.
Almost all of the participants ($29/30$) had dealt with privacy challenges in their work activities; they had also discussed and influenced privacy requirements of their systems.
Only one participant reported never being involved in privacy-related discussions, even though the systems handled sensitive personal data.
The participants ($29/30$) generally described a relatively informal privacy design process, not following any systematic approach for the requirements engineering process.
\begin{quote}
    \textit{``I guess it's informal. A general mindset is like, the less we have collected, the less we have to protect. I want to go to sleep at night, knowing that my software is secure. [...] I don't want to feel nervous about, you know, [that] the data mine that we're sitting on could be compromised.''} (P02).
\end{quote}

Similarly, most participants ($22/30$) reported \textbf{planning for privacy} when creating or changing systems, but without formal procedures or methodologies.
Privacy is typically considered an ongoing process of identifying and solving concerns, with a case-by-case analysis depending on the project.
Only four participants reported using a \textbf{privacy-by-design} approach in their organisations, but mostly without any clear methodologies.
Only one participant reported using Hoepman's Privacy Design Strategies \cite{hoepman2014privacy} from the Little Blue Book for Privacy \cite{hoepman2018privacy}.
These findings identify a knowledge gap in the practitioners' understanding of privacy engineering and its translation to the existing practices in the industry.
\\

\noindent\fbox{
    \begin{minipage}{.95\linewidth}
        \setlength\parskip{1em}
        \textit{\textbf{Synthetic findings:}} Privacy is generally addressed on a project basis without formal or systematic methods for eliciting requirements. Only a few participants reported using privacy-by-design principles and privacy design strategies to guide the software development activities.
    \end{minipage}
}

\subsubsection*{Sub-theme T3.2: Privacy Strategies}
Even though participants seldom articulate privacy by referring to the literature, regulations, and standards, they still reported several strategies used for addressing privacy concerns in their organisations.
Throughout the interviews, the participants were prompted to specify and discuss privacy strategies and controls they had experience with.
The list of well-known privacy strategies (shown in Figure \ref{fig:priveng-practices}), compiled by the study carried out by Hadar and colleagues \cite{hadar2018privacy}, was also posed to participants to ascertain whether they were familiar with or used any of the strategies.
The participants were further asked to bring up real-life examples if they used any strategies.
This list of privacy strategies should not be interpreted as a list of clear-cut categories but rather used to cover common high-level strategies, facilitating discussions with participants.

\begin{figure*}[h]
  \centering
  \includegraphics[width=1.0\linewidth]{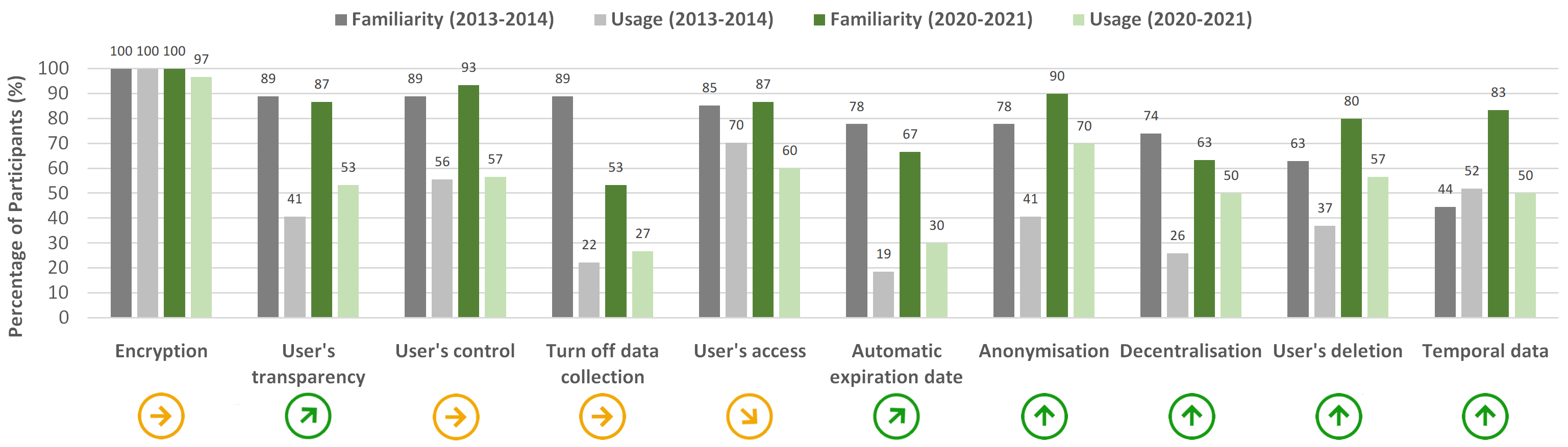}
  \caption{Percentages of participants' responses on the familiarity and usage of privacy strategies. The responses from the participants ($n=27$) in the study of Hadar \textit{et al.} (2018) \cite{hadar2018privacy} (in grey) can be compared to the responses from participants ($n=30$) in this study (in green).}
  \label{fig:privacy-strategies}
\end{figure*}

As shown in Figure \ref{fig:privacy-strategies}, most participants were familiar with and used most of these privacy strategies.
On average, they were familiar with 8/10 privacy strategies and used 5.5/10 of them.
A total of ten participants were familiar with all privacy strategies in the list, of which only 3 have used all of them in their careers.
Compared to the results in Hadar \textit{et al.}\cite{hadar2018privacy}, based on twenty-seven interviews carried out in 2013-2014, it is possible to see an overall upward trend in the adoption of privacy strategies among practitioners.
Especially for the privacy strategies of data anonymisation, decentralisation of datasets, user's data deletion, and regular deletion of personal data.
\textbf{Encryption} remains the most used privacy strategy, common to practitioners for protecting communication channels (e.g., HTTPS/SSL) and data storage (e.g., encrypted databases).

Only one participant reported in-depth knowledge of \textbf{anonymisation techniques}\footnote{During the interviews, the term \textbf{anonymisation} was broadly used to talk about anonymisation and other techniques (e.g., pseudo-anonymisation, de-identification, and obfuscation) with the participants. Nevertheless, anonymisation in its true sense has been increasingly proven to be infeasible since sufficiently motivated attackers can re-identify individuals or de-anonymise entire datasets \cite{ohm2009broken, zibuschka2019anonymization}, so anonymisation techniques turn out to be only capable of pseudo-anonymising the data.} (i.e., k-anonymity, t-closeness) used to anonymise medical data for governmental agencies.
Anonymisation was also used to present metrics in business reports (e.g., aggregated in Key Performance Indicators).
However, most participants ($16/30$) refer to anonymisation but actually employ pseudo-anonymisation techniques ($8/30$) or simply removing of identifiers ($8/30$).
The most common was relying on user IDs, e.g., universally unique identifiers (UUIDs) and advertising industry standard unique identifiers (ADIDs), which provide some level of obfuscation but can still be used by the platforms to link data of unique individuals as well as target them.
Notably, one participant questioned the practical use of such user IDs and privacy impacts.
\begin{quote}
    \textit{``I don't know the name and last name because it's decoupled. Sure. But with the user ID, I can target you on my platform, and they can show you what they want. So it can collect your interest associated with your user ID and show you advertisements based on your user ID. And my effect is real in the real world, even though I don't know you [...]''} (P19).
\end{quote}

The \textbf{decentralisation of datasets} is also becoming more common.
Practitioners ($6/30$) reported three types of decentralisation: (a) separation of types of data amongst different datasets ($2/30$); (b) separate databases for development and production environments ($1/30$); and (c) separate or partial access to databases for microservices and APIs ($3/30$).
However, two practitioners reported that their organisations are actually going the opposite way, investing in data lakes to centralise all data instead of decentralising it.

Data deletion strategies, such as \textbf{regularly deleting personal data} (i.e., temporal data) and \textbf{enabling users to delete their data}, have also been increasingly adopted.
Data is usually deleted to minimise storage costs, not because of privacy concerns.
Anyhow, practitioners ($16/30$) mentioned deleting transactional data (``data in-use'') and promptly deleting it after use, as well as deleting logs and analytics data of inactive users.
Also, apart from organisations with legal obligations to keep the data, most practitioners ($18/30$) mentioned that users could delete their data from the systems, sometimes referring to it as the right to be forgotten.
\begin{quote}
    \textit{``Yeah, that's something that kind of touches what I work on. [...] when the user decides to delete something, it's something that the company takes really seriously. So, of course, it's impossible to immediately delete something from everywhere. And so, that kind of triggers some flags that go all over the place, and the data kind of fades away with some time.''} (P25).
\end{quote}

Measures to \textbf{provide information to users about their data in the systems} (i.e., user's transparency) also show a slight upward trend in use.
Many practitioners ($7/30$) rely on their systems' privacy policy and user consent pages to provide such information.
However, many participants ($12/30$) were not involved with such parts of the system and did not know the extent of transparency provided to users.
In a similar slightly upward trend, \textbf{automatic expiration of personal data} is also being used, but even though most practitioners ($20/30$) alleged being familiar, only some ($9/30$) actually use it nowadays.
Here, practitioners ($5/30$) would usually refer to measures that are not necessarily specific to personal data, such as purging data regularly or using a data life cycle management\footnote{The term \textbf{data life cycle management} was used in the context of data governance, referring to an organisational approach for managing data throughout its entire life cycle, from data collection to data destruction.} strategy in the organisation.

Mechanisms that give users the \textbf{option to turn off data collection} are also barely used ($8/30$). 
Also, in such situations, some practitioners ($3/30$) actually refer to letting users revoke permissions or partially withdrawing consent in the system and not really specifying timeframes to turn off data collection.

A slightly downward trend was found for solutions that \textbf{enable users to access and view their data} as well as download a copy of it.
However, participants ($11/30$) reported solutions that give users only partial access to their data, such as user profile information, history of activity, etc.
Systems that allow users to access all the information gathered about them were reported just by two participants.
Also, only two participants mentioned functionalities that let users easily download all their data at once.
This shows that even basic privacy strategies, i.e., for data access requests, are still not sufficiently implemented.

The use of solutions that enable \textbf{user's control over privacy settings and what is revealed to other users or system operators} also has not changed significantly.
Although many participants ($15/30$) described specific systems that allow users to configure their privacy preferences, there are still some problematic reports, e.g., \textit{``in our system, the user doesn't have any control about their privacy settings''} (P14).
In one situation, the issue was even more complicated when handling employees' data.
\begin{quote}
    \textit{``the owner of the data is not the user, [it] is the customer. And that's transparent to everybody. Like my user knows the boss is looking right. So they know the data is not theirs.''} (P24)
\end{quote}

Besides the well-known strategies, the participants were asked at different points about their experience with privacy, e.g., when talking about previous projects, organisation procedures, design approaches, etc.
Table \ref{tab:other-strategies} lists many other privacy strategies participants have used in their work projects.
The strategies were classified into one of two areas (a) privacy and (b) security, even though they often overlap.
Security primarily refers to integrity and confidentiality, common pillars for informational privacy.
However, privacy has other dimensions, such as transparency, purpose limitation, data minimisation, accuracy, storage limitation, and accountability.

Noticeably, on the privacy side, many practitioners ($13/30$) have been involved in designing consent management mechanisms, either developing the user interfaces or the implications in the back end.
Having a ``data minimisation mindset'' was also stressed by many participants ($10/30$) in order to refrain from collecting unneeded data and avoid any privacy impacts altogether.
Only a few participants ($8/30$) also reported close contact with writing or requesting privacy policy changes to ensure users' transparency and compliance with regulations.
Even less prevalent ($5/30$) is the use of company-wide data classification strategies, such as data catalogues, that help enforce purpose limitation and data minimisation principles.

On the security side, apart from using encryption, traditional security controls for authentication and authorisation are still the most prevalent ($12/30$) data safeguarding methods, especially for controlling and logging access to personal data within organisations.
Other strategies, not primarily for addressing privacy and security issues, were occasionally reported, such as code review meetings ($3/30$) and data lifecycle procedures ($3/30$).
\\

\noindent\fbox{
    \begin{minipage}{.95\linewidth}
        \setlength\parskip{1em}
        \textit{\textbf{Synthetic findings:}} Participants were more aware of privacy strategies, compared to prior works \cite{hadar2018privacy, peixoto2020understanding, arizon2021understanding}, especially regarding anonymisation techniques. However, only encryption mechanisms are extensively adopted, suggesting that other privacy controls might have gained less traction over the years, even with several privacy laws being enacted.
    \end{minipage}
}

\begin{table}[htbp]
\caption{Other privacy and security strategies reported by the participants.}
\begin{center}\small
\begin{tabular}{|p{0.7\linewidth}cc|}
\hline
\textbf{Privacy strategies} & \textbf{T/O} & \textbf{Freq.}\\
\hline
User consent mechanisms & T & 13 \\
Minimise collection of personal data & O & 10 \\
Writing and updating privacy policies & O & 8 \\
Data classification and data catalogue tools & T & 5 \\
Data sovereignty and cross-border data transfer procedures & O & 4 \\
Checklists to guide following regulations & O & 4 \\
Agreements for data sharing with third-parties & O & 3 \\
Data steering committee & O & 3 \\
Data retention policies & O & 2 \\
Privacy Impact Assessments & O & 2 \\
\hline
\hline
\textbf{Security strategies} & \textbf{T/O} & \textbf{Freq.}\\
\hline
User authentication and access control & T & 12 \\
Developers access control to databases & T & 12 \\
Incident management plan and procedures & O & 6 \\
Logging developers' access to personal data & T & 4 \\
APIs and microservices limited access to databases & T & 4 \\
Certifications for products and services (e.g., ISO, Common Criteria) & O & 2 \\
Penetration testing procedures & T & 2 \\
Physical access control to network space & T & 2 \\
\hline
\hline
\textbf{Other strategies} & \textbf{T/O} & \textbf{Freq.}\\
\hline
Code review meetings & O & 3 \\
Data life cycle management procedures & O & 3 \\
\hline
\end{tabular}
\label{tab:other-strategies}
\end{center}
\textbf{Notes:} Technical (\textbf{T}) or Organisational (\textbf{O}) control types. The frequencies (\textbf{Freq.}) are merely informative.
\end{table}

\section{Discussion}
\label{sec:discussion}
This study provides a thematic framework that highlights the interactions between three core themes of the practitioner's mindset and stance, organisational privacy aspects, and privacy engineering practices.
Different challenges and opportunities arise from this triple interaction, with implications for academic researchers and industry practitioners.
Hence, in the following subsections, we discuss the key identified challenges and pathways for future research.

\subsection{Instilling Personal Privacy Values}
Privacy engineering is a team effort involving multiple internal and external stakeholders working together to develop highly secure and privacy-preserving systems.
Among the main conclusions, this work shows that practitioners are becoming more knowledgeable about privacy concerns, laws and regulations.
We also noticed a positive change of mindset among the participants, perhaps driven by the new wave of privacy regulations, demanding them to adjust their engineering practices and underlying business models.
This study suggests that practitioners perceive privacy tasks with more priority, as opposed to the earlier findings from \cite{hadar2018privacy} and \cite{arizon2021understanding}.
Examples of that were practitioners expressing a personal sense of responsibility for privacy, a duty as a developer, and often raising privacy issues, which superiors and other team members welcomed.

Nonetheless, many practitioners also view privacy as a significant burden.
This myth that ``privacy is hard'' has been challenged by researchers in privacy engineering \cite{hoepman2021privacy}.
Likewise, we argue that privacy needs to be accounted for in the entire software development life cycle, from the conceptual design to the decommissioning of a system.
If one leaves privacy to the end as an ``add-on'' to the system, it becomes much harder, if not impossible, to implement it.
Software engineers should think of privacy the same way mechanical engineers think regarding safety when designing cars.
Safety requirements demand more knowledge, planning and resources throughout the design, but it is a non-negotiable: people need safe cars; an unsafe car is not good enough.
In the digital society that we live in today, people need privacy and software systems should effectively support it.

Academia can contribute to the education of future generations of software practitioners by approaching topics such as ethics and instilling responsible design and innovation as core tenets of software development.
In industry, practitioners ought to be able to raise issues and challenge bad engineering decisions; the connivance or complicity with privacy violations is unacceptable.
However, to make it work, such personal privacy values, which could be developed through training and education, must also align with the organisational values.

\subsection{Researching Organisational Privacy Culture and Climate}
Organisational privacy culture is a powerful force that can lead to positive or negative privacy climates, influencing the practitioners' attitudes and behaviours, which is in line with the previous findings from \cite{hadar2018privacy, sirur2018are, bednar2019engineering, ribak2019translating, tahaei2021privacy, arizon2021understanding}.
Here it is clear that organisations need to build a privacy climate that encourages open communication of privacy concerns among practitioners.
This study also shows evidence that the commitment to privacy from people in leadership and supervisory positions is crucial to facilitate conversations within and across teams in the organisations.
Examples are the team leaders, superiors and ``disguised privacy champions'' that play an essential role in helping to identify and solve privacy problems.

Organisations should improve privacy knowledge by aiming for changes in attitudes, e.g., using meaningful privacy training and awareness programs, thus creating the desire to change behaviour.
As discussed in \cite{arizon2021understanding}, desired privacy behaviours also need to be rewarded in the organisations, and misalignments between top management and software developers regarding responsibilities and the importance of privacy need to be addressed.
Here we can draw an analogy with Microsoft's Secure Development Lifecycle (SDL), a security model that consists of a series of security-focused activities and deliverables linked to each phase of the software development process \cite{lipner2004trustworthy}.
As an organisation-wide approach, the SDL also provoked a shift in the organisational security culture, showing the company's commitment to protecting customers and building trust.
Such models can also be created for privacy, taking advantage of the current advances in privacy engineering.

Apart from development teams, many other departments in the organisations also bear responsibility for privacy and data protection (e.g., human resources, finances, marketing, sales and administration) since they also may deal with large amounts of personal data daily.
Professionals in all departments may face privacy issues when handling the personal data of users, customers and employees, so there are many open fronts for investigation.
Future work in such areas is essential, especially to understand cross-departmental interactions, tensions and challenges that can affect practitioners involved with privacy-sensitive systems.
One example is the work from Degli Esposito \cite{degli2016dataveillance} also discusses the creation and effects of organisational privacy culture, though focusing on the interplay of legal privacy regimes and big data analytics in companies.
Da Veiga and Martins \cite{daveiga2015information} have also initially introduced an information protection culture assessment (IPCA) instrument for measuring aspects of privacy culture across the organisation.

Currently, the field of Organisational Privacy Culture and Climate (OPCC) remains largely unexplored from information systems and software engineering perspectives.
A scoping review on the topic has shown that the area is still emerging and that more primary research is needed to substantiate its theory, constructs, and practical instruments \cite{iwaya2022organisational}.
Future research is also needed for clarifying distinctions and integrating theories for the two constructs of ``organisational privacy climate'' and ``organisational privacy culture'', as previously discussed in section \ref{sec:org-climate-culture}.
Studies such as \cite{daveiga2015information}, \cite{hadar2018privacy}, and \cite{arizon2021understanding} are some examples of quantitative and qualitative research in the OPCC area.
As mentioned, the present study has corroborated and presented new findings, but replication studies are still fundamental for future work.
The body of knowledge of OPCC, in fact, still deals with an over-reliance on WEIRD populations.
Similarly, privacy engineering as an emerging field also has limitations, as already highlighted in user privacy studies \cite{saleh2019habibi, naveed2022ask}.

\subsection{Evaluating Privacy Engineering Practices}
A fresher view of the concrete privacy strategies and practices that software practitioners are adopting was also one of the motivations for this research. 
This investigation was partially possible due to the crystallisation of privacy engineering as a research field and the further development of privacy standards and guidelines.
On this account, evidence still shows a rather unstructured and inconsistent uptake of privacy strategies, procedures, controls and standards.
For instance, primary methodologies for privacy engineering, such as Privacy Impact Assessments (PIAs), are rarely used in practice.

The overall unsystematic approach to privacy suggests that complex tasks of privacy risk analysis and threat modelling are mostly done informally (e.g., brought up in meetings and discussions) by practitioners that may have limited knowledge and experience concerning privacy.
Less frequently, participants in large enterprises mentioned dedicated privacy and security departments, internal privacy policies, training and awareness programs that made them very confident in their companies' current privacy practices.

Even so, many well-known privacy engineering methods and techniques (e.g., PIAs, threat modelling, privacy design strategies, and privacy design patterns) have received little attention from practitioners.
Researchers also have some responsibility here in bridging the schism between such privacy engineering artefacts and actual software engineering practices \cite{kostova2020privacy}, in order to propose industry-informed and feasible methodologies that practitioners can realistically use.
Existing solutions for privacy engineering also need to be carefully evaluated to determine their effectiveness, feasibility, and reliability in practice.
For instance, some well-known methodologies for privacy risk analysis, i.e., LINDDUN \cite{deng2011privacy} and PIAs, are also known for being time-consuming and hard to scale, especially if we think in terms of service-orientated architectures (hundreds of services).

Some researchers suggest that the automation of some of the repetitive steps in such privacy engineering techniques could turn them into standard everyday practices \cite{zibuschka2019analysis, zimmermann2020automation}.
Automated testing tools can be used to check the security configuration of servers and communication channels and that the personal data is being collected according to specified purposes, helping organisations achieve ``continuous privacy compliance'' \cite{li2022towards}.
However, such approaches still have to be further developed and evaluated.

\section{Threats to Validity}
\label{sec:threats-validity}
This study has a few limitations that should be considered when interpreting the findings.
Here, we refer to these limitations as threats to validity using the classification proposed by Runeson and H\"{o}st \cite{runeson2009guidelines} on the study's construct validity, internal and external validity, and reliability.

\emph{Construct validity} refers to the congruence of the operational measures (i.e., instruments) with what the researchers have in mind and what is investigated according to the research questions \cite{runeson2009guidelines}.
The main instruments used in this study are provided as supplementary materials in ``Appendix A -- Screening Survey'' and ``Appendix B -- Interview Guide''.
New questions were added to the interview guide to emphasise the uptake of privacy practices.
Two external researchers evaluated these instruments, and later we piloted the instruments in one interview.
After reviewing and adjusting the instruments, all the authors were satisfied with the screening survey and interview guide.

\emph{Internal validity} is concerned with the examination of causal relationships, in particular, if other factors that are not being examined in the study can affect the results \cite{runeson2009guidelines}.
It is worth noting that interviewees are not real-life observations, meaning that participants can only talk retrospectively based on their memory and experiences during a relatively short time (e.g., one hour).
In such positions, participants tend to emphasise the topic being discussed, so behaviours might appear more prominent than the actual importance and priority that are given to real-life circumstances.
However, to mitigate this issue, when we asked questions to the participants we further asked them to bring real-life cases and situations to clarify and make explicit the reported privacy behaviours.
Follow-up questions were also frequently used to determine the depth of knowledge and practices reported by the participants.

\emph{External validity} refers to the extent that findings can be generalised \cite{runeson2009guidelines}.
We do not claim the generalisability of our findings since the 30 participants from 29 different organisations cannot provide a statistically representative sample of the entire population of software practitioners.
The participants with varied backgrounds were recruited, many from outside European and North American regions, but it should be noted that they are still mostly from Western countries.
The views of practitioners from other regions shall be considered in future research.
Nonetheless, we were able to draw similar findings comparable to prior work on the topic, such as \cite{hadar2018privacy, peixoto2020understanding, arizon2021understanding}.

\emph{Reliability} relates to the extent to which the data and analysis depend on the specific researchers \cite{runeson2009guidelines}.
Even though the entire research group has been involved in this project from the beginning, the first author mostly led and performed the thematic analysis.
To mitigate this issue, an external research fellow with expertise in qualitative research on security and privacy contributed to the initial data analysis steps.
Two researchers used an open coding strategy and later merged the preliminary codebooks.
In addition, periodic reviews among the research team were used to seek agreement during the qualitative coding process, discussing the generated codes and the creation of the themes and the final thematic framework. 
All authors have security, privacy, software engineering, and qualitative research expertise.

\section{Conclusion}
\label{sec:conclusion}
In this study, we aimed to investigate the topic of information privacy and its engineering approaches through the perspectives of software practitioners.
The results of this study are valuable to multiple stakeholders, such as (1) practitioners who are establishing and improving privacy practices in their organisations and teams; (2) researchers who want to understand the current state-of-the-practice on privacy, and get insights for building more industry-informed privacy engineering theories, methods, techniques and tools; (3) policy-makers who can understand the practitioners' views and adoption of existing regulations, as well as their needs for clearer regulatory guidelines and supporting documents; and, (4) standardisation bodies who are interested in knowing the current uptake of privacy standards and norms by the industry globally.

Software practitioners worldwide are increasingly more conscious of information privacy when developing software systems.
Most practitioners know existing privacy regulations and are familiar with many common privacy strategies, e.g., encryption, transparency, data access and control, and anonymisation.
However, there is still a significant gap between research and practice regarding privacy engineering.
Even though many technical standards and methodologies around privacy engineering have already been proposed, practitioners need to be made aware of such initiatives.
This study also shows that developers generally have an unstructured and inconsistent set of approaches for handling and solving privacy issues.

Our findings also highlight the importance of creating and fostering a privacy culture in the organisation.
Practitioners' privacy behaviours are strongly influenced by their environment, so corporate values should match the principles enshrined in the existing privacy laws.
In order to create a conducive environment, the top management needs to demonstrate commitment to privacy values, aligning themselves to the employees' values and beliefs, communicating their importance, and rewarding desired privacy behaviours.
This interplay of research areas, such as privacy engineering, organisational privacy culture, and socio-technical systems, remains largely unexplored, offering several pathways for future research in terms of qualitative and quantitative research.

In future work, we plan on further investigating and building the theory around organisational privacy culture and climate, given its strong influence on practitioners.
More qualitative research is needed to identify the primary constructs, as well as quantitative research for creating practical instruments to assess and measure privacy climate.
We also intend to conduct further studies to evaluate the existing privacy engineering methods and techniques and assess their feasibility and practicality.
Besides that, we also see the need for more focused interview-based studies in under-researched application areas, e.g., with practitioners developing social networks and advertising platforms, in which negative privacy impacts can be amplified.

\ifCLASSOPTIONcompsoc
  \section*{Acknowledgments}
\else
  \section*{Acknowledgment}
\fi

The work has been supported by the Cyber Security Research Centre Limited whose activities are partially funded by the Australian Government's Cooperative Research Centres Programme.
The work of L.H. Iwaya has also been partially supported by the European Commission’s H2020 Programme via the CyberSec4Europe project (Grant: 830929), the Swedish Knowledge Foundation via the TRUEdig project, and the Region V\"{a}rmland through the DHINO project (Grant: RUN/220266) and by Vinnova via the DigitalWell Arena project (Grant: 2018-03025).
Awais Rashid is supported by REPHRAIN, National Research Centre on Privacy, Harm Reduction and Adversarial Influence Online (EPSRC Grant: EP/V011189/1).
The authors would also like to thank Dr. Chamila Wijayarathna for his invaluable support on the research design, data collection, and the first steps of data analysis.
We also thank our research fellows Roshan Rajapakse and Nesara Dissanayake for reviewing and providing extensive feedback on all the instruments used in this study.

\ifCLASSOPTIONcaptionsoff
  \newpage
\fi



%

\bibliography{myrefs} 
\bibliographystyle{ieeetr}

%

\begin{IEEEbiography}[{\includegraphics[width=1in,height=1.25in,clip,keepaspectratio]{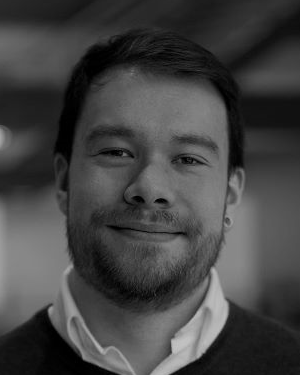}}]{Leonardo Horn Iwaya} is an Associate Senior Lecturer in the Department of Mathematics and Computer Science, Karlstad University, Sweden. He obtained a Ph.D. degree in computer science from Karlstad University, Sweden. He also holds a M.Sc. degree in electrical engineering from the University of S\~{a}o Paulo and a B.Sc. degree in computer science from Santa Catarina State University, Brazil. From 2011 to 2014, he was a Research Assistant with the Laboratory of Computer Networks and Architecture (LARC), PCS-EPUSP. From 2019 to 2021, he worked as a Postdoctoral Researcher with the School of Computer Science, The University of Adelaide, Australia, as part of the Cyber Security Cooperative Research Centre (CSCRC). During 2021 and 2022, he worked as a Postdoctoral Researcher with the Interdisciplinary Research Group on Knowledge, Learning and Organizational Memory (KLOM), Federal University of Santa Catarina. Currently, he works with the Privacy \& Security (PriSec) Research Group, Karlstad University, contributing to projects, such as CyberSecurity4Europe, TRUEdig, SURPRISE, and DigitalWell Arena. His research interests include privacy engineering, information security, human factors, mobile and ubiquitous health systems, and the privacy impacts of new technologies.
\end{IEEEbiography}


\begin{IEEEbiography}[{\includegraphics[width=1in,height=1.25in,clip,keepaspectratio]{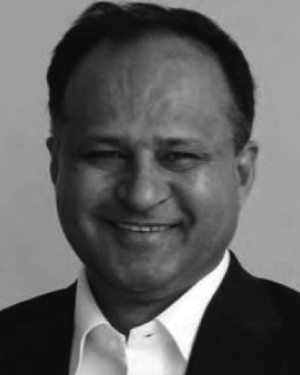}}]{M. Ali Babar} M. Ali Babar is a Professor in the School of Computer Science, University of Adelaide, Australia. He leads a theme on architecture and platform for security as service in Cyber Security Cooperative Research Centre (CSCRC), a large initiative funded by the Australian government, industry, and research institutes. Professor Babar is the technical project lead of one of the largest projects on “Software Security” in ANZEC region funded by the CSCRC. SOCRATES brings more than 75 researchers and practitioners from 6 research providers and 4 industry partners for developing and evaluating novel knowledge and AI-based platforms, methods, and tools for software security. After joining the University of Adelaide, Prof Babar established an interdisciplinary research centre called CREST, Centre for Research on Engineering Software Technologies, where he directs the research, development and education activities of more than 25 researchers and engineers in the areas of Software Systems Engineering, Security and Privacy, and Social Computing. Professor Babar’s research team draws a significant amount of funding and in-kind resources from governmental and industrial organisations. Professor Babar has authored/co-authored more than 275 peer-reviewed research papers at premier Software journals and conferences. Professor Babar obtained a Ph.D. in Computer Science and Engineering from the school of computer science and engineering of University of New South Wales, Australia. He also holds a M.Sc. degree in Computing Sciences from University of Technology, Sydney, Australia. More information on Professor Babar can be found at http://malibabar.wordpress.com.
\end{IEEEbiography}

\begin{IEEEbiography}[{\includegraphics[width=1in,height=1.25in,clip,keepaspectratio]{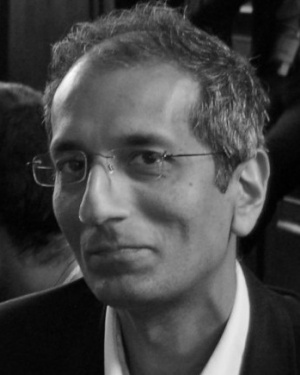}}]{Awais Rashid} is a Professor of Cyber Security at the University of Bristol, UK. His research focuses on security and privacy in large connected infrastructures. He is Director of REPHRAIN, the National Research Centre on Privacy, Harm Reduction and Adversarial Influence Online. He is also Director of the EPSRC Centre for Doctoral Training in Trust, Identity, Privacy and Security in Large-scale Infrastructures. He also heads a major international effort on developing a Cyber Security Body of Knowledge (CyBOK) and leads projects as part of the UK Research Institute on Trustworthy Industrial Control Systems (RITICS), UK Research Institute on Socio-technical Cyber Security (RISCS), the National Centre of Excellence on Cyber Security of Internet of Things (PETRAS) and the ESRC Hub+ on Digital Security by Design (Discribe). He held a Fellowship of the Alan Turing Institute and, prior to joining the University of Bristol, was co-founder and co-director of the Security Lancaster Institute at Lancaster University.\end{IEEEbiography}




\end{document}